\documentclass{article}
\usepackage{PRIMEarxiv}
\usepackage[utf8]{inputenc} 
\usepackage[T1]{fontenc}    
\usepackage{hyperref}       
\usepackage{url}            
\usepackage{booktabs}       
\usepackage{amsfonts}       
\usepackage{nicefrac}       
\usepackage{microtype}      
\usepackage{lipsum}
\usepackage{fancyhdr}       
\usepackage{graphicx}       
\graphicspath{{media/}}     

\usepackage{moreverb}
\usepackage{graphicx}

\usepackage[utf8]{inputenc}
\usepackage{tikz}
\usepackage{booktabs}
\usepackage{rotating}
\usepackage{amsmath}
\usepackage{amsthm}
\usepackage{amssymb}
\usepackage{tablefootnote}
\usepackage{float}
\usepackage{lscape}
\usepackage{caption}
\usepackage{chngcntr}
\usepackage{algorithm}
\usepackage{algpseudocode}
\usepackage{dsfont}
\usepackage{graphicx}
\usepackage{dirtytalk}
\usepackage{appendix}
\usepackage{setspace}
\usepackage{titlesec}

\setdisplayskipstretch{1}
\titlespacing{\section}{0pt}{0.5\baselineskip}{0.5\baselineskip}
\titlespacing{\subsection}{0pt}{0.5\baselineskip}{0.5\baselineskip}

\newcommand\independent{\protect\mathpalette{\protect\independenT}{\perp}}
\def\independenT#1#2{\mathrel{\rlap{$#1#2$}\mkern2mu{#1#2}}}

\title{Covariate Adjustment in Bayesian Adaptive Randomized Controlled Trials}
\author{
  James Willard, Shirin Golchi, and Erica EM Moodie \\
  Department of Epidemiology and Biostatistics \\
  McGill University \\
  Montreal, Canada
}

\begin{document}
\maketitle
\vspace{-5mm}

\begin{abstract}
In conventional randomized controlled trials, adjustment for baseline values of covariates known to be at least moderately associated with the outcome increases the power of the trial.  Recent work has shown particular benefit for more flexible frequentist designs, such as information adaptive and adaptive multi-arm designs.  However, covariate adjustment has not been characterized within the more flexible Bayesian adaptive designs, despite their growing popularity. We focus on a subclass of these which allow for early stopping at an interim analysis given evidence of treatment superiority. We consider both collapsible and non-collapsible estimands, and show how to obtain posterior samples of marginal estimands from adjusted analyses. We describe several estimands for three common outcome types. We perform a simulation study to assess the impact of covariate adjustment using a variety of adjustment models in several different scenarios. This is followed by a real world application of the compared approaches to a COVID-19 trial with a binary endpoint. For all scenarios, it is shown that covariate adjustment increases power and the probability of stopping the trials early, and decreases the expected sample sizes as compared to unadjusted analyses. 
\end{abstract}

\keywords{Bayesian adaptive designs \and covariate adjustment \and clinical trials \and power \and stopping criteria}

\section{Introduction}

In conventional, fixed size randomized controlled trials (RCTs), adjustment for baseline values of covariates known to be at least moderately associated with the outcome has been shown to increase the power of the trial \cite{hernandez_2004_cov_adj, hernandez_2006_tte_adj, benkeser2021improving, harrell2021bbr}. This is because covariate adjustment improves the precision of the estimated treatment effect and accounts for the outcome heterogeneity within each treatment arm that is explained by the adjustment variables \cite{ich1999harmonised, senn2013seven, benkeser2021improving}.  Adjustment also corrects for any chance imbalance of important baseline variables which may exist post-randomization \cite{kahan2014risks}. Therefore, covariate adjustment in the primary analysis of clinical trials is now recommended by both the US Food and Drug Administration (FDA) and European Medicines Agency (EMA) \cite{food148910adjusting, european2015guideline}.  Additionally, systematic reviews have suggested its use in practice has grown over time \cite{austin2010substantial, ciolino2019ideal}.  Recently, these power increases have been demonstrated in more flexible frequentist designs, such as information adaptive and adaptive multi-arm designs \cite{van2022combining, lee2022benefits}.  However, the simulation scenarios investigated in each of these designs contained at least one of the following: only continuous outcomes with no treatment-covariate interactions where the marginal and conditional estimands are the same; only a single sample size; a data generating process containing only a small number of variables; or only a small number of covariate adjustment models. A more comprehensive investigation is needed to better understand the benefits of covariate adjustment under a broader array of flexible design scenarios.

While the impact of covariate adjustment has been demonstrated in the flexible frequentist designs mentioned above, it has not been characterized within flexible Bayesian adaptive designs, where early stopping at an interim analysis is permitted given evidence of treatment superiority or futility. In these designs one may learn about a treatment effect while potentially requiring fewer participants than fixed designs. Additionally, Bayesian trials allow for seamless incorporation of prior information for model parameters including covariate effects. The impact of combining prior information with covariate adjustment has not been previously investigated.  With the growing interest in Bayesian adaptive designs, a characterization of covariate adjustment for several commonly used sample sizes and outcome types would be highly valuable for researchers.

In this work, we consider the impact of covariate adjustment in Bayesian adaptive designs which allow for early stopping for superiority and provide a step-by-step tutorial for post adjustment marginalization. We explore several data generating processes for continuous, binary, and time-to-event outcomes, and consider adjustment models which include several forms of misspecification while incorporating varying levels of prior information for the covariate effects.  The covariate adjustment described herein is performed using generalized linear models (GLMs). However, the methods, results, and recommendations discussed below are not specific to these models, and are expected to generalize to other parametric and nonparametric models.

The manuscript is organized in the following manner. We first introduce and describe Bayesian adaptive designs which allow for early stopping, as well as the targets of inference (i.e., estimands), which are marginal treatment effects for each endpoint. We then describe the specific collapsible and non-collapsible estimands used in this manuscript. For the non-collapsible estimands, we describe their estimation and marginalization through a Bayesian framework. This is followed by a simulation study which shows the impact of covariate adjustment on design operating characteristics including power, the probability of stopping the trial early, and expected sample size, for multiple sample sizes. We then show a real-world application of covariate adjustment within a COVID-19 RCT and end with a discussion.

\section{Bayesian Adaptive Designs with Early Stopping}\label{bgsd}
Bayesian adaptive designs allow for predetermined changes to the trial design at interim analyses based on evidence provided by the accumulating data \cite{berry2010bayesian, giovagnoli2021bayesian}. These designs include sequentially randomized trials which allow for early stopping for superiority or futility at interim analyses. Interim analyses are performed to determine whether to stop the trial early and declare treatment superiority or futility, or to continue the trial. The decision to stop the trial early at an interim analysis is controlled by a predefined decision rule. Decision criteria may be defined with respect to different functionals of the posterior distribution of the parameter of interest or estimand. Posterior or posterior predictive probability statements about the estimand are commonly used statistics.  In the RCT setting, estimands are typically defined as marginal treatment effects in a specified population of interest.  We adopt this convention throughout, but delay further discussion of marginal estimands and their posterior estimation until the next section. 

In the Bayesian adaptive designs described in this work, interim or final decisions are defined with respect to posterior probability statements about a marginal treatment effect,  $\gamma(\boldsymbol{\theta})$, which is a function of model parameters $\boldsymbol{\theta}$. The alternative hypothesis of the trial is formulated as this marginal treatment effect being greater than a clinically meaningful threshold, $\gamma_0$:
\begin{equation}
    H_0: \gamma(\boldsymbol{\theta}) \le \gamma_0\ \text{vs}\ H_A: \gamma(\boldsymbol{\theta}) > \gamma_0. \nonumber
\end{equation}
\noindent A Bayesian test statistic can be defined as the posterior probability of this alternative hypothesis given the data, $\mathbf{\mathcal{D}}_{n_t} = \{Y_{n_t}, A_{n_t}, \mathbf{X}_{n_t \times p}\}$, which may include any observed outcomes ($Y_{n_t}$), treatment assignments ($A_{n_t}$), and $p$ additional covariates ($\mathbf{X}_{n_t \times p}$), for the $n_t$ participants who are enrolled in the trial at an interim or final analysis conducted at time $t$:
\begin{equation}\label{test_stat}
T(\mathbf{\mathcal{D}}_{n_t})=P(H_A| \mathbf{\mathcal{D}}_{n_t})=P(\gamma(\boldsymbol{\theta}) > \gamma_0 | \mathbf{\mathcal{D}}_{n_t}).
\end{equation}
\noindent This statistic is then used to define a decision rule, which declares treatment superiority at any interim or final analysis if the statistic exceeds some upper probability threshold, $u$, i.e., if
\begin{equation}
    T(\mathbf{\mathcal{D}}_{n_t}) > u. \nonumber
\end{equation}
\noindent If a superiority declaration is made at an interim analysis, the trial stops early. The common approach in the design of Bayesian adaptive trials is the \say{hybrid} approach, where the upper probability threshold, $u$, is optimized so that the trial design has desirable values of frequentist operating characteristics \cite{berry2010bayesian}. For example, power ($\mathcal{P}$) and the Type 1 error rate (T1E) are defined as follows:
\begin{equation}
    \begin{aligned}
    \mathcal{P} &= P(T(\mathbf{\mathcal{D}}_{n_t}) > u \mid \gamma(\boldsymbol{\theta})=\gamma^* > \gamma_0) \\
    \text{T1E} &= P(T(\mathbf{\mathcal{D}}_{n_t}) > u \mid \gamma(\boldsymbol{\theta}) = \gamma_0). \nonumber
\end{aligned}
\end{equation}
\noindent Since the sampling distribution of a Bayesian posterior probability is generally unknown, calibration of the design to meet frequentist operating characteristics requires simulation studies. Note that models that adjust for covariates under most settings result in analytically intractable posterior distributions. Therefore, the evaluation of $T(D_{n_t})$ within every trial simulation requires posterior sampling or approximation techniques. In this paper we use Markov chain Monte Carlo (MCMC) to sample from the posterior distribution when not available in closed form.

\section{Estimands and Bayesian Estimation}\label{marg_procedures}

In what follows, let $A_i$ be defined as a binary treatment assignment for the $i^{th}$ participant, where $A_i=1$ represents being randomized to the treatment group and $A_i=0$ to the control group. Let $\widetilde{\mathbf{X}}_{n_t \times d}$ represent a matrix of $j=1,...,d$ covariates measured at baseline for $i=1,...,n_t$ participants in the study at an interim or final analysis conducted at time $t$.  Let $\tilde{\mathbf{x}}_i=(x_1, x_2,...,x_d)_i$ represent the row of this matrix corresponding to the full covariate pattern of the $i^{th}$ participant, and let $Y_i$ be an arbitrary outcome of interest for the $i^{th}$ participant. For notational simplicity, subscripts are dropped for the remainder of this article except when strictly necessary. Treatments are assigned through simple randomization and are thus independent of all covariates measured at baseline (i.e.,~$A \independent \tilde{\mathbf{x}})$, but we allow for the possibility of chance imbalance of covariates between treatment groups.

We use the term \say{unadjusted analysis} to refer to a model which includes only the treatment assignment indicator ($A$), and the term \say{adjusted analysis} for a model which includes $p \le d$ additional covariates, $\mathbf{X} \subseteq \widetilde{\mathbf{X}}$, with $\mathbf{x}_i=(x_1,...,x_p)_i$ representing the row of $\mathbf{X}$ corresponding to the $i^{th}$ participant's covariate pattern used for adjustment.  For adjusted analyses, many different covariate sets may be adjusted for in addition to the binary treatment indicator.  This includes any covariates $\mathbf{Z} \subseteq \mathbf{X}$ where a treatment-covariate interaction exists. Let $\phi$ be the regression coefficient for the treatment assignment indicator, $\beta_0$ be the intercept, $\boldsymbol{\beta} = \{\beta_1,...,\beta_p\}$ be the vector of covariate main effects, and $\boldsymbol{\omega} = \{\omega_1,...,\omega_m\}$ be the vector of treatment-covariate interaction effects for those covariates used in the adjustment model within a GLM setting. Additionally, let $\boldsymbol{\zeta}$ be a set of nuisance parameters not of direct interest but which are required for model specification (e.g., baseline hazard parameters in a time-to-event setting). Then let $\boldsymbol{\theta} = \{\beta_0, \phi, \boldsymbol{\beta}, \boldsymbol{\omega}, \boldsymbol{\zeta}\}$ be the set of model parameters with prior $p(\boldsymbol{\theta})$, and define $\eta(\boldsymbol{\theta}; A_i,  \mathbf{X}_i, \mathbf{Z}_i)$ to be the expected outcome for participant $i$ on the linear scale.  Let $p(Y_i \mid A_i, \mathbf{X}_i, \boldsymbol{\theta})$ represent each participant's contribution to the likelihood function and $g(\cdot)$ be the link function. Assuming independence among participants, we have the joint posterior distribution of the model parameters, $\pi(\boldsymbol{\theta} \mid \mathcal{D}_{n_t})$, under arbitrary adjustment model specification, being proportional to the likelihood times the prior:
\begin{equation}
    \begin{aligned}
    \pi(\boldsymbol{\theta} \mid \mathcal{D}_{n_t}) &\propto \prod_{i=1}^{n_t} p(Y_i \mid A_i, \mathbf{X}_i, \boldsymbol{\theta})p(\boldsymbol{\theta}). \nonumber\\
    \end{aligned}
\end{equation}
In this manuscript, the decision rule used at an interim or final analysis is defined with respect to a posterior probability statement about a marginal estimand $\gamma(\boldsymbol{\theta})$, presented in Equation \ref{test_stat}, which is the average treatment effect in a population of interest. We note this marginal estimand can be defined as a contrast of population average quantities, such as a difference of means or a ratio of population-level event or survival probabilities. Let $\mu(\boldsymbol{\theta};A)$ be the population level parameter for treatment group $A$ used as input for contrast $f(\cdot)$. Then $\gamma(\boldsymbol{\theta})$ can be represented by:
\begin{equation}
    \gamma(\boldsymbol{\theta}) = f(\mu(\boldsymbol{\theta};A=1), \mu(\boldsymbol{\theta};A=0)). \nonumber
\end{equation}
Unadjusted analyses yield posterior samples from the marginal parameter $\mu(\boldsymbol{\theta};A)$ directly:
\begin{equation}
    \mu(\boldsymbol{\theta};A) = g^{-1}(\eta(\boldsymbol{\theta};A)). \nonumber
\end{equation}
Adjusted analyses, however, yield posterior samples from the conditional parameter $\mu(\boldsymbol{\theta};A, \mathbf{X})$ for fixed covariate pattern $\mathbf{X}$:
\begin{equation}
    \mu(\boldsymbol{\theta};A, \mathbf{X}) = g^{-1}(\eta(\boldsymbol{\theta};A, \mathbf{X}, \mathbf{Z})). \nonumber
\end{equation}
\noindent When a treatment effect is \textit{collapsible}, it can be represented as a contrast between either the marginal or conditional parameters for a fixed covariate pattern $\mathbf{X}$:
\begin{equation}
\begin{aligned}
    \gamma(\boldsymbol{\theta}) &= f(\mu(\boldsymbol{\theta};A=1), \mu(\boldsymbol{\theta};A=0)) \\
    &= f(\mu(\boldsymbol{\theta};A=1,\mathbf{X}), \mu(\boldsymbol{\theta};A=0,\mathbf{X})). \nonumber
\end{aligned}
\end{equation}
\noindent Thus, under collapsibility, samples from the posterior distribution of the marginal estimand can be obtained from either an unadjusted or adjusted analysis. When a treatment effect is \textit{non-collapsible}, the marginal treatment effect cannot be represented as a contrast between conditional parameters for a fixed covariate pattern $\mathbf{X}$ \cite{daniel2021making}. This is commonly the case for treatment effects modeled by GLMs or in the presence of treatment-covariate interactions. As an example, consider a hypothetical RCT with a binary endpoint which follows the following logistic regression model with binary treatment assignment $A$ and binary covariate $X$, where $P(X=1)=P(A=1)=0.5$ and where $\phi=\log(5)$, $\beta=\log(10)$, and $\boldsymbol{\theta}=\{\phi,\beta\}$:
\begin{equation}
    \text{logit}(P(Y=1 \mid A, X)) = \phi A + \beta X. \nonumber
\end{equation}
Define $\mu(\boldsymbol{\theta};A,X)$ to be the treatment specific conditional risk. The conditional odds ratio for those who are treated versus untreated can be represented as a contrast of these conditional risk parameters, and is 5 regardless of the value of $X$.  To find the marginal odds ratio, the treatment specific conditional risks must be averaged with respect to the distribution of $X$ before calculating the odds ratio, effectively collapsing over $X$ in a stratified two-by-two table. Doing so gives treatment specific marginal risks $\mu(\boldsymbol{\theta};A)$ which are used to obtain a marginal odds ratio of 4.1. This value is smaller than that for the conditional odds ratio and shows the marginal odds ratio cannot be represented as a contrast of the treatment specific conditional risks. Thus, the odds ratio is non-collapsible (see Section G of the Online Supplementary Material for more details). Under non-collapsibility, samples from the posterior distribution of the marginal estimand can still be obtained directly from unadjusted analyses, but not from adjusted analyses. To obtain samples from the posterior distribution of the marginal estimand using an adjusted analysis, the posterior samples of $\mu(\boldsymbol{\theta};A,\mathbf{X})$ must be marginalized with respect to the distribution of $\mathbf{X}$, yielding samples of $\mu(\boldsymbol{\theta};A)$ which are then used in the contrast:
\begin{equation}
\mu(\boldsymbol{\theta};A) = \int_{\mathbf{X}}\mu(\boldsymbol{\theta};A, \mathbf{X})p(\mathbf{X})d\mathbf{X}. \nonumber
\end{equation}
\noindent This marginalization is commonly called standardization or Bayesian G-computation for point treatments \cite{kalton1968standardization, freeman1980summary, lane1982analysis, saarela2015predictive, remiro2020marginalization, keil2018bayesian, daniel2021making}. The integral over $p(\mathbf{X})$ is approximated through summation where, for each of $s=1,...,S$ Monte Carlo samples $\boldsymbol{\theta}_s$ from $\pi(\boldsymbol{\theta} \mid \mathcal{D}_{n_t})$, the following calculations is performed:
\begin{equation}
\mu(\boldsymbol{\theta}_s; A) \approx \sum_{i=1}^{n_t} w_{i,s} \mu(\boldsymbol{\theta}_s; A, \mathbf{x}_i). \label{standardize_mu} 
\end{equation}
\noindent The $\mathbf{x}_i$ are the covariate patterns used for adjustment and which are contained in the joint empirical distribution of the collected sample data.  The weights $\mathbf{w}_s=(w_{1,s},...,w_{n_t,s})$ are sampled as $\mathbf{w}_s \sim \text{Dirichlet}(\mathbf{1}_{n_t})$, corresponding to the Bayesian bootstrap, where $\mathbf{1}_{n_t}$ is the $n_t$-dimensional vector of $1$'s \cite{rubin1981bayesian, oganisian2021practical, linero2023and}. This can then be used to obtain a single sample from the posterior of the marginal treatment effect:
\begin{equation}
    \gamma(\boldsymbol{\theta}_s) = f(\mu(\boldsymbol{\theta}_s;A=1), \mu(\boldsymbol{\theta}_s;A=0)). \nonumber
\end{equation}
\noindent Note that when any $X_j$ for $j=1,...,p$ is the propensity score being jointly modeled with the outcome of interest, a different Bayesian bootstrap procedure should be utilized \cite{stephens2022causal}. For the remainder of the manuscript, it is assumed $\mathbf{X}$ does not contain a propensity score estimated in this manner.  
 
The procedure above enables $S$ samples from the posterior distribution of the marginal estimand $\gamma(\boldsymbol{\theta})$ to be obtained using an arbitrary adjustment model.  This allows for a direct performance comparison between adjustment models within Bayesian adaptive designs.  Below we describe this procedure within the context of specific models that are most commonly used for different outcome types in clinical trials.  

\subsection{Collapsible Treatment Effects}
\subsubsection{Difference in Means: No Treatment-Covariate Interactions}
Consider the difference in means of a continuous endpoint under the assumption of no treatment-covariate interactions (i.e., $\mathbf{Z}=\emptyset$; homogeneity of the treatment effect),
\begin{equation}
    \gamma(\boldsymbol{\theta}) := \mu(\boldsymbol{\theta};A=1) - \mu(\boldsymbol{\theta};A=0) \nonumber
\end{equation}
\noindent where $\mu(\boldsymbol{\theta};A=a) = E[Y \mid A=a;\boldsymbol{\theta}]$. This marginal estimand represents the difference in expected outcomes between those assigned to treatment versus those assigned to control. Estimation proceeds assuming independent outcomes and the following model:
\begin{equation}
    \begin{aligned}
        &p(Y_i \mid A_i,\mathbf{X}_i, \boldsymbol{\theta}) = \text{Normal}(\mu(\boldsymbol{\theta};A_i,\mathbf{X}_i),\sigma^2) \\
        &\mu(\boldsymbol{\theta};A_i,\mathbf{X}_i) = \beta_0 + \phi A_i + \mathbf{X}_i\boldsymbol{\beta} \\
        &\boldsymbol{\theta} = \{\beta_0, \phi, \boldsymbol{\beta}, \sigma^2\}. \nonumber
    \end{aligned}
\end{equation}
\noindent Since this treatment effect is collapsible, posterior samples of the marginal estimand can be obtained from either an unadjusted or adjusted analysis, and samples from the posterior distribution of the treatment indicator coefficient $\phi$ are commonly used.

\subsection{Non-collapsible Treatment Effects}
\subsubsection{Difference in Means: Treatment-Covariate Interactions}
Consider again a continuous outcome, but under the assumption of at least one treatment-covariate interaction (i.e., $\mathbf{Z} \ne \emptyset$; treatment effect heterogeneity).  The difference in means is non-collapsible. Estimation proceeds assuming independent outcomes and the following model:  
\begin{equation}
    \begin{aligned}
        &p(Y_i \mid A_i,\mathbf{X}_i, \boldsymbol{\theta}) = \text{Normal}(\mu(\boldsymbol{\theta};A_i,\mathbf{X}_i),\sigma^2) \\
        &\mu(\boldsymbol{\theta};A_i,\mathbf{X}_i) = \beta_0 + \phi A_i + \mathbf{X_i}\boldsymbol{\beta} + (A_i \cdot \mathbf{Z_i})\boldsymbol{\omega} \\
        &\boldsymbol{\theta} = \{\beta_0, \phi, \boldsymbol{\beta}, \boldsymbol{\omega}, \sigma^2\}. \nonumber
    \end{aligned}
\end{equation}
\noindent Since this estimand is non-collapsible, posterior samples of the conditional $\mu(\boldsymbol{\theta};A,\mathbf{X})$ must be marginalized using (\ref{standardize_mu}) before forming the contrast to obtain a posterior sample of the marginal difference in means. An outline of this procedure is provided in Appendix A in the Online Supplementary Materials.

\subsubsection{Relative Risk and Odds Ratio} \label{relative_risk}
Consider a dichotomous outcome which is modelled as a Bernoulli random variable.  Examples of commonly used marginal estimands include the relative risk
\begin{equation}
\gamma(\boldsymbol{\theta}) := \mu(\boldsymbol{\theta};A=1)/\mu(\boldsymbol{\theta};A=0) \nonumber
\end{equation}
\noindent and the odds ratio
\begin{equation}
\gamma(\boldsymbol{\theta}) := \frac{\mu(\boldsymbol{\theta};A=1)/(1-\mu(\boldsymbol{\theta};A=1))}{\mu(\boldsymbol{\theta};A=0)/(1-\mu(\boldsymbol{\theta};A=0))} \nonumber
\end{equation}
\noindent where $\mu(\boldsymbol{\theta};A=a)=E[Y\mid A=a;\boldsymbol{\theta}]$. The relative risk represents the ratio comparing the risk of an event for those assigned to treatment versus those assigned to control. The odds ratio represents the ratio comparing the odds of an event for those assigned to treatment versus those assigned to control.  Estimation proceeds assuming independent outcomes and the following model:  
\begin{equation}
\begin{aligned}
        &p(Y_i \mid A_i,\mathbf{X}_i, \boldsymbol{\theta}) = \text{Bernoulli}(\mu(\boldsymbol{\theta};A_i,\mathbf{X}_i)) \\
        &\mu(\boldsymbol{\theta};A_i,\mathbf{X}_i) =  \text{logit}^{-1}(\beta_0 + \phi A_i + \mathbf{X}_i\boldsymbol{\beta} + (A_i \cdot \mathbf{Z}_i)\boldsymbol{\omega}) \\
        &\boldsymbol{\theta}=\{\beta_0, \phi, \boldsymbol{\beta}, \boldsymbol{\omega}\}. \nonumber
\end{aligned}
\end{equation}
To obtain posterior samples of the marginal relative risk or odds ratio, posterior samples of the conditional $\mu(\boldsymbol{\theta};A,\mathbf{X})$ must be marginalized using (\ref{standardize_mu}) before forming the contrast, an outline of  which is provided in Appendix A in the Online Supplementary Materials.

\subsubsection{Hazard Ratio}\label{tte_trt_measure}
Let $T$ denote the time to an event of interest. Let $h(t\mid A)$ represent the hazard, the instantaneous event rate at time $t$, for those assigned to treatment $A$:
\begin{equation}
    h(t \mid A)= \lim_{\Delta t \to 0} \frac{P(t \le T < t + \Delta t \mid T>t, A)}{\Delta t}. \nonumber
\end{equation}
Under the assumption of no competing risks, there is a one-to-one relationship between the hazard and survival probability at time $t$:
\begin{equation}
    \begin{aligned}
    S(t \mid A)&=\exp\left(-\int_{v=0}^{t}h(v \mid A)dv\right). \nonumber \\
    \end{aligned}
\end{equation}
\noindent Further assuming proportional hazards, an estimand of interest is the marginal hazard ratio
\begin{equation}
\gamma(\boldsymbol{\theta}) := \log\{\mu(\boldsymbol{\theta};A=1)\}/\log\{\mu(\boldsymbol{\theta};A=0)\} \nonumber
\end{equation}
\noindent where $\mu(\boldsymbol{\theta};A=a)=S(t\mid A=a;\boldsymbol{\theta})$.  This estimand represents the ratio comparing the hazard of those assigned to treatment versus those assigned to control. We note that the estimation framework described below is general and may be utilized to target other estimands of interest (e.g., the risk difference or risk ratio).

Under an RCT framework which allows for right censoring only, the time from a participant's initial enrollment to an event of interest may occur after the trial has ended.  Let $T_i$ be the $i^{th}$ participant's observed event time or right censoring time. Let $\delta_i$ be the $i^{th}$ participant's observation indicator, where $\delta_i=1$ means the event time is observed before the end of the trial and where $\delta_i=0$ means the event time is right censored. Let $Y_i=\{T_i, \delta_i\}$ be the observed data. On the hazard scale, the hazard of the event at time $t$ for the $i^{th}$ participant can be modeled as below, where $h_0(t)$ is the baseline hazard function:
\begin{equation}
\begin{aligned}
h_i(t \mid A_i, \mathbf{X}_i)&=h_0(t)\exp(\eta_i) \\
\eta_i &= \phi A_i + \mathbf{X}_i \boldsymbol{\beta} + (A_i \cdot \mathbf{Z_i})\boldsymbol{\omega}.
\nonumber
\end{aligned}
\end{equation}
\noindent This yields the corresponding survival probability:
\begin{equation}
\begin{aligned}
    S_i(t \mid A_i, \mathbf{X}_i)&=\exp\left(-\int_{v=0}^{t}h_i(v \mid A_i, \mathbf{X}_i)dv\right) \\
    &=\exp\left(-\int_{v=0}^{t}h_0(v)\exp(\eta_i)dv\right). \nonumber \\
\end{aligned} 
\end{equation}
The baseline hazard function may be flexibly modeled, with one possible choice being through M-splines \cite{brilleman2020bayesian}. Let $M(t;\boldsymbol{\psi}, \boldsymbol{k},\delta)$ be an M-spline function:
\begin{equation}
    M(t;\boldsymbol{\psi}, \boldsymbol{k},\delta)=\sum_{l=1}^L \psi_lM_l(t;\boldsymbol{k},\delta). \nonumber
\end{equation}
\noindent Here $\boldsymbol{\psi}$ is the vector of coefficients for the $L$ M-spline basis terms, with degree $\delta$ and knot locations $\boldsymbol{k}$. Integrating this M-spline function yields the following I-spline function, which is evaluated using the same coefficients, degree and knot locations:
\begin{equation}
    I(t;\boldsymbol{\psi}, \boldsymbol{k},\delta)=\sum_{l=1}^L \psi_lI_l(t;\boldsymbol{k},\delta). \nonumber
\end{equation}
\noindent Both M-spline and I-spline functions can be evaluated analytically \cite{splines2}. By flexibly modeling the baseline hazard with M-splines, the hazard and the survival probability become, respectively:
\begin{equation}
    \begin{aligned}
    h_i(t \mid A_i, \mathbf{X}_i)&=M(t;\boldsymbol{\psi}, \boldsymbol{k},\delta)\exp(\eta_i) \\
    S_i(t \mid A_i, \mathbf{X}_i)&=\exp\left(-\int_{v=0}^{t}M(v;\boldsymbol{\psi}, \boldsymbol{k},\delta)\exp(\eta_i)dv\right) \\
    &=\exp\left(-I(t;\boldsymbol{\psi}, \boldsymbol{k},\delta)\exp(\eta_i)\right). \nonumber \\
    \end{aligned}
\end{equation}
\noindent Estimation then proceeds by assuming independent outcomes and the following model:  
\begin{equation}
\begin{aligned}
    &p(Y_i \mid A_i, \mathbf{X}_i, \boldsymbol{\theta}) = S_i(T_i \mid A_i, \mathbf{X}_i)^{1-\delta_i}h_i(T_i \mid A_i, \mathbf{X}_i)^{\delta_i} \\ 
    &\boldsymbol{\theta}=\{\boldsymbol{\psi}, \phi, \boldsymbol{\beta}, \boldsymbol{\omega}\}. \nonumber
\end{aligned} 
\end{equation}
Posterior samples of the marginal hazard ratio are obtained by first marginalizing samples of $\mu(\boldsymbol{\theta};A,\mathbf{X})={S(t\mid A=a,\mathbf{X};\boldsymbol{\theta})}$ using (\ref{standardize_mu}) and then forming the contrast \cite{daniels2023bayesian, stitelman2011targeted, remiro2020marginalization}. This procedure is  outlined in Appendix A in the Online Supplementary Materials.

\section{Simulation Study}\label{sim_study}

In this section we perform simulations for the design and models described in the previous sections. We consider a design with a superiority stopping rule where superiority is declared at any interim or final analysis performed at time $t$ if $T(\mathcal{D}_{n_t}) > 0.99$. The same value of $u=0.99$ is selected for all maximum sample sizes to control the overall Type 1 error rate of the unadjusted model (i.e., Type 1 error rate below 0.05). The unadjusted model is selected as a conservative choice for trial planning purposes, since the true strength of any covariate effects and adjustment benefit may not be known in practice at the trial planning stage \cite{benkeser2021improving}. Note that our interest is in comparing the performance of different adjustment models, not optimizing $u$ for each maximum sample size and model, so a single conservative value of $u$ above is chosen for all simulations. Our marginal treatment effects of interest are the difference in means of a Normal endpoint under the assumption of no treatment-covariate interactions, the relative risk under a binary endpoint, and the hazard ratio under a time-to-event endpoint. Data generating processes with five covariates and a treatment assignment indicator are used for multiple sample sizes with each endpoint. We consider adjustment models which include several forms of misspecification and which incorporate varying levels of prior information for the covariate effects. To obtain marginal estimates from the adjustment models, the procedures described in the previous section are utilized.  We follow a setup similar to that reported in a previous study which investigated covariate adjustment for endpoints commonly used in COVID-19 RCTs \cite{benkeser2021improving}: for each maximum sample size with each endpoint, three treatment effect values are chosen.  The first is the null treatment effect, and the second and third are those where the unadjusted model achieves roughly 50\% and 80\% power. This excludes the scenarios of a maximum sample size of 100 under the binary and time-to-event endpoints, whose second and third treatment effect sizes are chosen as those where the unadjusted model achieves roughly 30\% or 40\% power. This ensures all simulations maintain realistic values for the marginal treatment effects, and that the impact of covariate adjustment is compared at a value of the treatment effect for which trials are commonly powered (i.e., 80\%). For each maximum sample size with each endpoint, the impact of covariate adjustment is quantified through the values of the following design operating characteristics: power, Type 1 error rate, expected sample size, probability of stopping early, bias and root mean squared error (RMSE). 

\subsection{Data Generating Mechanisms}\label{cov_distn}

For each combination of endpoint and maximum sample size $\{100, 200, 500, 1000\}$, the data generating mechanisms for the treatment assignment and covariate distributions measured at baseline are shown below, where joint independence between all variables is assumed.  Letting $\eta$ represent the linear predictor used in the data generating mechanisms, the set $\{\boldsymbol{\beta}, \phi\}$ represents the conditional covariate and treatment effects on the linear predictor scale:
 \begin{equation}
    \begin{aligned}
    &\eta = \beta_0 + \phi A+\beta_1X_1 + \beta_2X_2 + \beta_3X_3 +\beta_4X_3^2 + \beta_5X_5 \\
    &\{A, X_1, X_2, X_6\} \sim \text{Bernoulli}(0.5) \\
    &\{X_3, X_5, X_7, X_8\}\sim \text{Normal}(0,1) \\
        &\boldsymbol{\beta}=  (\beta_0, \beta_1, \beta_2, \beta_3, \beta_4, \beta_5)  \notag \\
        &\gamma_{\text{max ss}}=\mathcal{S}(\eta) \\
        &\boldsymbol{\Theta} = \{(\gamma_{100} \times \boldsymbol{\beta}), (\gamma_{200} \times \boldsymbol{\beta}), (\gamma_{500} \times \boldsymbol{\beta}), (\gamma_{1000} \times \boldsymbol{\beta})\}. \nonumber \\
        \end{aligned}
\end{equation}
\noindent Note that the variables $\{X_6,X_7,X_8\}$ are noise and are not predictive of, or correlated with, any other variables in the data generating mechanism.  For the binary endpoint, $\beta_0$ is optimized to generate datasets which exhibit the correct marginal control event risk of $p_{ctr}=0.3$ (Appendix B in the Online Supplementary Materials). For the continuous and time-to-event endpoints, $\beta_0=0$. For the time-to-event endpoint, an exponential baseline hazard with rate $\lambda=0.02$ is used.  For the non-collapsible treatment effects, the true values of the marginal estimand $\gamma=\gamma(\boldsymbol{\theta})=f(\mu(\boldsymbol{\theta};A=1), \mu(\boldsymbol{\theta};A=0))$ do not equal the conditional treatment effects $\phi=f(\mu(\boldsymbol{\theta};A=1,\mathbf{X}), \mu(\boldsymbol{\theta};A=0,\mathbf{X}))$ for fixed $\mathbf{X}$.  Thus, the reported values of the marginal estimands are obtained through simulation (denoted by $\mathcal{S}(\cdot)$; Appendix B in the Online Supplementary Materials), and the values of $\gamma$ and $\phi$ are reported together. For the continuous endpoint, $\boldsymbol{\beta}=(0, 0.5, -0.25, 0.5, -0.05, 0.25)$. For the binary endpoint, $\boldsymbol{\beta}=(-1.26, 1, -0.5, 1, -0.1, 0.5)$. For the time-to-event endpoint, $\boldsymbol{\beta}=(0, 1, -0.5, 1, -0.1, 0.5)$. For each maximum sample size (max ss) within each outcome type, 1,000 treatment-covariate datasets are generated. These are used to generate 1,000 different outcome vectors for each value of the marginal treatment effect within the corresponding maximum sample size.  The specific parameter values used for all simulations are included in Table \ref{tbl:sim_study_pars}.

\begin{table}
    \caption{Simulation study parameter settings for the marginal estimand ($\gamma$) and conditional treatment effect ($\phi$) for each endpoint and maximum sample size (max ss).  The marginal estimands for the continuous, binary, and time-to-event are, respectively, the difference in means under the assumption of no treatment-covariate interactions, the relative risk, and the hazard ratio.}
    \centering
   \begin{tabular}{ccc}
    \toprule
      Continuous & Binary & Time-to-event  \\
      \midrule \\
     {\scriptsize \begin{tabular}{r r | r r} 
         \toprule
         max ss & $\gamma$ & max ss & $\gamma$\\ [0.5ex]
         \midrule
         100 & 0 & 500 & 0 \\ 
         100 & -0.52 & 500 & -0.22 \\ 
         100 & -0.73 & 500 & -0.32 \\ 
         200 & 0 & 1000 & 0 \\
         200 & -0.36 & 1000 & -0.16 \\
         200 & -0.52 & 1000 & -0.22 \\ [1ex] 
         \bottomrule
        \end{tabular} }
        & {\scriptsize\begin{tabular}{r r r | r r r} 
         \toprule
         max ss & $\gamma$ & $\phi$ & max ss & $\gamma$ & $\phi$\\ [0.5ex]
         \midrule
         100 & 1 & 0 & 500 & 1 & 0 \\ 
         100 & 0.53 & -0.99 & 500 & 0.72 & -0.56\\ 
         100 & 0.46 & -1.21 & 500 & 0.60 & -0.82\\ 
         200 & 1 & 0 & 1000 & 1 & 0 \\
         200 & 0.59 & -0.86 & 1000 & 0.80 & -0.39 \\
         200 & 0.41 & -1.36 & 1000 & 0.72 & -0.54 \\ [1ex] 
         \bottomrule
        \end{tabular} }
        & {\scriptsize\begin{tabular}{r r r | r r r} 
         \toprule
         max ss & $\gamma$ & $\phi$ & max ss & $\gamma$ & $\phi$\\ [0.5ex]
         \midrule
         100 & 1 & 0 & 500 & 1 & 0 \\ 
         100 & 0.65 & -0.68 & 500 & 0.78 & -0.39\\ 
         100 & 0.60 & -0.79 & 500 & 0.71 & -0.54\\ 
         200 & 1 & 0 & 1000 & 1 & 0 \\
         200 & 0.69 & -0.59 & 1000 & 0.85 & -0.27 \\
         200 & 0.57 & -0.86 & 1000 & 0.78 & -0.39 \\ [1ex] 
         \bottomrule
        \end{tabular}}\nonumber \\ \\ \bottomrule
    \end{tabular}
    \label{tbl:sim_study_pars}
\end{table}

\subsection{Adjustment Models}
  
\noindent Six adjustment models are considered for all endpoints:
\begin{enumerate}
    \item correct: $\beta_0 + \phi A+\beta_1X_1 +\beta_2X_2 +\beta_3X_3 +\beta_4X_3^2 +\beta_5X_5$ 
    \item no quad: $\beta_0 +  \phi A+\beta_1X_1 +\beta_2X_2 +\beta_3X_3 +\beta_5X_5$ 
    \item correct noise: $\beta_0 +  \phi A+\beta_1X_1 +\beta_2X_2 +\beta_3X_3 +\beta_4X_3^2 +\beta_5X_5 +\beta_6X_6 +\beta_7X_7 +\beta_8X_8$ 
    \item correct prior: $\beta_0 + \phi A+\beta_1^{\dagger}X_1 +\beta_2^{\dagger}X_2 +\beta_3^{\dagger}X_3 +\beta_4^{\dagger}X_3^2 +\beta_5^{\dagger}X_5$ 
    \item correct strong prior: $\beta_0 + \phi A+\beta_1^{\dagger\dagger}X_1 +\beta_2^{\dagger\dagger}X_2 +\beta_3^{\dagger\dagger}X_3 +\beta_4^{\dagger\dagger}X_3^2 +\beta_5^{\dagger\dagger}X_5$ 
    \item unadjusted: $\beta_0 +  \phi A$. 
\end{enumerate}
The \textit{correct} model corresponds to an adjustment model which matches the data generating mechanism. The \textit{no quad} model drops the quadratic component of $X_3$ from the \textit{correct} model.  The \textit{correct noise} model adds noise variables $\{X_6,X_7,X_8\}$ to the \textit{correct} model. These three models include priors for all parameters which are weakly informative only. The \textit{correct prior} model is the same as the \textit{correct} model, but includes priors for the covariate effects centered at the values used in the data generating mechanism.  Similarly, the \textit{correct strong prior} model both centers and re-scales these priors to be more informative. Note that the prior for the treatment indicator coefficient remains weakly informative in these models.  Finally, the \textit{unadjusted} model includes only the binary treatment indicator and uses weakly informative priors. 


\begin{figure}
    \includegraphics[width=1\textwidth]{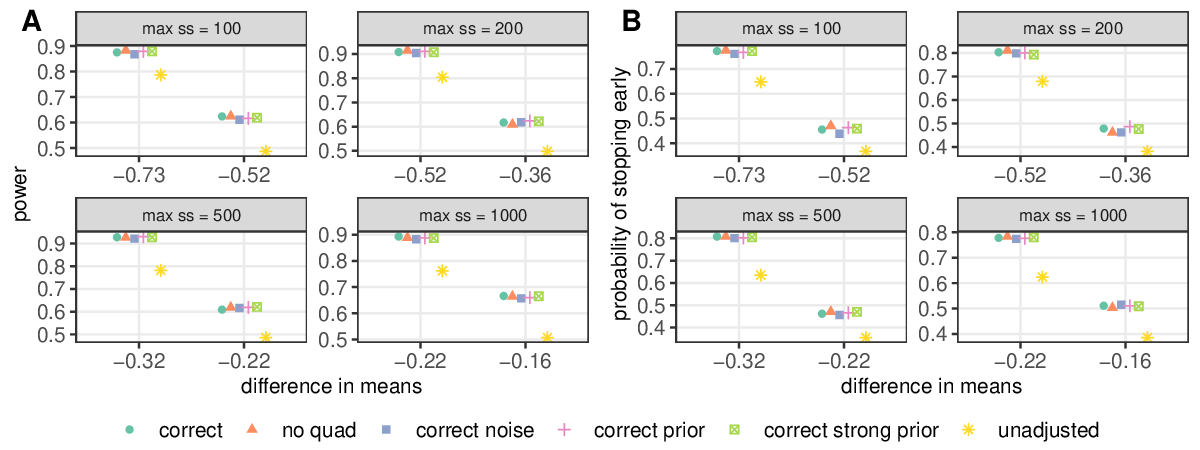}
    \caption{Continuous outcome. A) Power and B) probability of stopping early. Panels correspond to various maximum sample sizes (max ss). Points are jittered horizontally.}
    \centering
    \label{fig:cont_pow_pse}
\end{figure}

\begin{table}
\centering
\caption{Continuous outcome. Type 1 error rate (T1E), bias under the null ($\text{Bias}^*$), and expected sample size at three different values of the marginal difference in means ($\gamma$).}
\resizebox{\linewidth}{!}{
\begin{tabular}[t]{lrrrrrrrrrr}
\toprule
\multicolumn{1}{c}{ } & \multicolumn{5}{c}{Maximum sample size = 100} & \multicolumn{5}{c}{Maximum sample size = 200} \\
\cmidrule(l{3pt}r{3pt}){2-6} \cmidrule(l{3pt}r{3pt}){7-11}
\multicolumn{3}{c}{ } & \multicolumn{3}{c}{Expected sample size} & \multicolumn{2}{c}{ } & \multicolumn{3}{c}{Expected sample size} \\
\cmidrule(l{3pt}r{3pt}){4-6} \cmidrule(l{3pt}r{3pt}){9-11}
Adjustment model & T1E & $\text{Bias}^*$ & $\gamma=0$ & $\gamma=-0.52$ & $\gamma=-0.73$ & T1E & $\text{Bias}^*$ & $\gamma=0$ & $\gamma=-0.36$ & $\gamma=-0.52$\\
\midrule
correct & 0.034 & -0.013 & 98.3 & 78.6 & 62.5 & 0.039 & -0.011 & 196.8 & 154.7 & 116.0\\
no quad & 0.037 & -0.015 & 98.2 & 78.0 & 62.5 & 0.032 & -0.009 & 197.5 & 156.2 & 115.4\\
correct noise & 0.031 & -0.012 & 98.6 & 80.3 & 64.8 & 0.036 & -0.012 & 196.9 & 156.2 & 118.7\\
correct prior & 0.032 & -0.012 & 98.5 & 78.3 & 62.9 & 0.043 & -0.012 & 196.4 & 154.5 & 116.3\\
correct strong prior & 0.033 & -0.012 & 98.3 & 78.4 & 62.5 & 0.041 & -0.012 & 196.6 & 155.0 & 116.8\\
unadjusted & 0.028 & -0.007 & 98.5 & 82.6 & 68.9 & 0.039 & -0.017 & 196.5 & 163.1 & 130.7\\
\addlinespace
\toprule
\multicolumn{1}{c}{ } & \multicolumn{5}{c}{Maximum sample size = 500} & \multicolumn{5}{c}{Maximum sample size = 1000} \\
\cmidrule(l{3pt}r{3pt}){2-6} \cmidrule(l{3pt}r{3pt}){7-11}
\multicolumn{3}{c}{ } & \multicolumn{3}{c}{Expected sample size} & \multicolumn{2}{c}{ } & \multicolumn{3}{c}{Expected sample size} \\
\cmidrule(l{3pt}r{3pt}){4-6} \cmidrule(l{3pt}r{3pt}){9-11}
Adjustment model & T1E & $\text{Bias}^*$ & $\gamma=0$ & $\gamma=-0.22$ & $\gamma=-0.32$ & T1E & $\text{Bias}^*$ & $\gamma=0$ & $\gamma=-0.16$ & $\gamma=-0.22$\\
\midrule
correct & 0.036 & -0.008 & 491.1 & 389.1 & 290.4 & 0.031 & -0.003 & 987.0 & 753.0 & 598.8\\
no quad & 0.031 & -0.007 & 493.0 & 388.8 & 291.4 & 0.028 & -0.003 & 987.8 & 755.2 & 597.5\\
correct noise & 0.032 & -0.006 & 492.8 & 391.1 & 293.0 & 0.030 & -0.003 & 987.8 & 753.2 & 599.0\\
correct prior & 0.032 & -0.007 & 492.8 & 388.5 & 293.8 & 0.031 & -0.003 & 987.2 & 753.5 & 598.0\\
correct strong prior & 0.032 & -0.007 & 492.8 & 388.8 & 291.0 & 0.029 & -0.003 & 986.8 & 754.8 & 599.0\\
unadjusted & 0.032 & -0.006 & 492.4 & 415.6 & 343.1 & 0.028 & -0.002 & 989.0 & 817.8 & 690.2\\
\bottomrule
\end{tabular}}
\label{tbl:cont_table}
\end{table}

All simulations are performed using R (version 4.2.1).  All modeling is performed using the GLM and survival functionality of the rstanarm package (version 2.21.2), a front-end to the STAN probabilistic programming language \cite{r_proj, rstanarm, brilleman2020bayesian}. For all coefficients other than $\{\beta_j^{\dagger}, \beta_j^{\dagger\dagger}\}$, the package's default weakly informative priors are used \cite{rstanarm_priors}.  These priors induce moderate regularization and help improve computational stability. The prior for the intercept is placed after all covariates have been internally centered by rstanarm. Under a Normal likelihood for the continuous endpoint, this equates to the following priors, where $\beta_{0,c}$ represents the intercept's prior after covariate centering has been performed: 
\begin{equation}
\begin{split}
    \begin{aligned}
        \beta_{0,c} &\sim \text{Normal}(\bar{y}, 2.5s_y) \\
        \phi ~ &\sim \text{Normal}(0, 2.5(s_y/s_x)) \\
        \beta_j ~ &\sim \text{Normal}(0, 2.5(s_y/s_x)) 
    \end{aligned}
\end{split}
\hspace{10mm}
\begin{split}
\begin{aligned}
        \beta_j^{\dagger} ~ &\sim \text{Normal}(\beta_j, 2.5(s_y/s_x)) \\
        \beta_j^{\dagger\dagger} ~ &\sim \text{Normal}(\beta_j, s_y/s_x) \\
        \sigma &\sim \text{Exponential}(1/s_y).
        \nonumber
    \end{aligned}
\end{split}
\end{equation}

\noindent Under the binary and time-to-event endpoints, the above priors for $\{\phi, \boldsymbol{\beta}\}$ are used with $\bar{y}=0$ and $s_y=1$.  For the time-to-event endpoint, the coefficients of the M-spline basis ($\boldsymbol{\psi}$) are constrained to a simplex to ensure identifiability of both the basis and linear predictor intercepts.  Thus, the basis coefficients receive rstanarm's default Dirichlet prior \cite{brilleman2020bayesian}. All models specify three Markov chains, each with 2,000 posterior samples.  Half of the samples within each chain are used during the warm-up period, so 3,000 posterior samples in total are available for inference. Given the scale of the simulations performed, visual diagnostics assessing convergence of the Markov chains are not performed.  Rather, for all simulations, values of STAN's implementation of the Gelman-Rubin $\hat{R}$ statistic are assessed to ensure Markov chain convergence \cite{rhat_gelman_rubin}. 

The following null and alternative hypotheses are specified for the continuous, binary and time-to-event (TTE) endpoints, where $\gamma(\boldsymbol{\theta})$ is the marginal difference in means, marginal relative risk, and marginal hazard ratio, respectively:
\begin{equation}
\begin{aligned}
    \text{Continuous: } &H_0: \gamma(\boldsymbol{\theta}) \ge 0\ \text{vs}\ H_A: \gamma(\boldsymbol{\theta}) < 0 \nonumber \\
    \text{Binary and TTE: } &H_0: \gamma(\boldsymbol{\theta}) \ge 1\ \text{vs}\ H_A: \gamma(\boldsymbol{\theta}) < 1. \nonumber
\end{aligned}  
\end{equation}
\noindent For the continuous and binary endpoints, all outcomes are assumed to be observed immediately upon participant enrollment. For the time-to-event endpoint, it is assumed all outcomes are observed strictly after enrollment.  For the continuous endpoint, interim analyses are performed after every 25, 50, 125, and 250 participants are enrolled for maximum samples sizes of 100, 200, 500, and 1,000, respectively. For the binary and time-to-event endpoints, interim analyses are event driven.  For the binary endpoint, interim analyses are performed after at least 10, 20, 50, and 100 new events occur for maximum sample sizes 100, 200, 500, and 1,000, respectively. For the time-to-event endpoint, interim analyses are performed after at least 20, 40, 100, and 200 new events occur for maximum sample sizes 100, 200, 500, and 1,000, respectively. These numbers are chosen for each endpoint to ensure that on average the total number of analyses performed under the null treatment effect is less than four, which helps control the Type 1 error rate. They are also large enough to ensure there is a moderate chance of stopping at an early interim analysis under the non-null treatment effects.  For the continuous and binary endpoints, interim analyses are performed until the trial is stopped early for superiority or until the maximum sample size is reached, at which time the final analysis is performed. For the time-to-event endpoint, interim analyses are performed until the trial is stopped early for superiority or until 50 time units from the start of the trial is reached, at which time the final analysis is performed. For this endpoint, participant enrollment is permitted until 25 time units.  This ensures that participants enrolled at later time points are under follow-up long enough to have a moderately high probability of experiencing the event before the end of the trial.  It also ensures that the maximum number of participants will be enrolled if there is not clear evidence of superiority at an early interim analysis. No loss to follow-up is assumed and administrative censoring of those still at risk is performed at 50 time units from the start of the trial. 

\begin{figure}
    \includegraphics[width=1\textwidth]{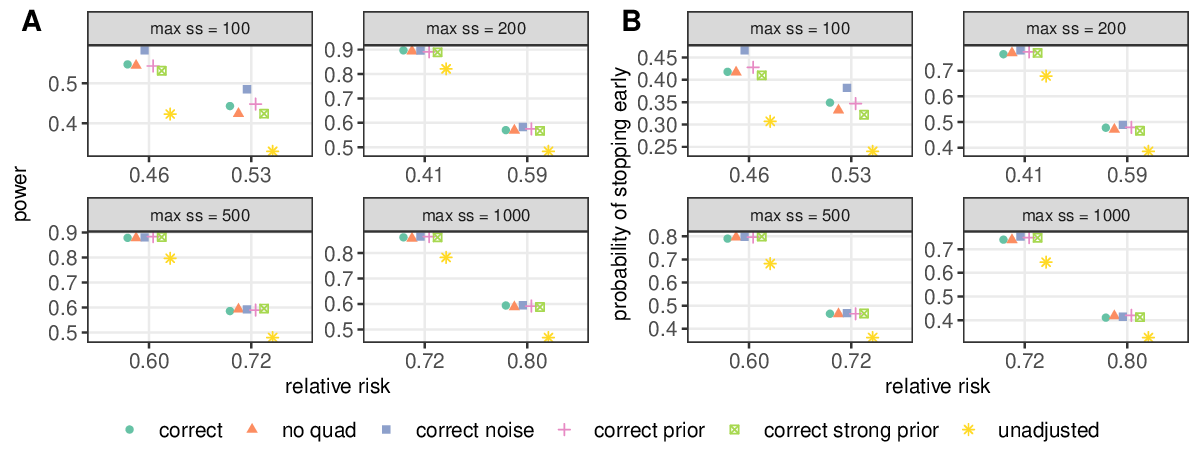}
    \caption{Binary outcome. A) Power and B) probability of stopping early. Panels correspond to various maximum sample sizes (max ss). Points are jittered horizontally.}
    \centering
    \label{fig:binary_pow_pse}
\end{figure}

\begin{table}
\centering
\caption{Binary outcome. Type 1 error rate (T1E), bias under the null ($\text{Bias}^*$), and expected sample size at three different values of the marginal relative risk ($\gamma$).}
\resizebox{\linewidth}{!}{
\begin{tabular}[t]{lrrrrrrrrrr}
\toprule
\multicolumn{1}{c}{ } & \multicolumn{5}{c}{Maximum sample size = 100} & \multicolumn{5}{c}{Maximum sample size = 200} \\
\cmidrule(l{3pt}r{3pt}){2-6} \cmidrule(l{3pt}r{3pt}){7-11}
\multicolumn{3}{c}{ } & \multicolumn{3}{c}{Expected sample size} & \multicolumn{2}{c}{ } & \multicolumn{3}{c}{Expected sample size} \\
\cmidrule(l{3pt}r{3pt}){4-6} \cmidrule(l{3pt}r{3pt}){9-11}
Adjustment model & T1E & $\text{Bias}^*$ & $\gamma=1$ & $\gamma=0.53$ & $\gamma=0.46$ & T1E & $\text{Bias}^*$ & $\gamma=1$ & $\gamma=0.59$ & $\gamma=0.41$ \\
\midrule
correct & 0.063 & 0.031 & 97.0 & 86.3 & 83.6 & 0.036 & 0.021 & 196.8 & 162.5 & 138.3\\
no quad & 0.053 & 0.035 & 97.4 & 86.9 & 83.7 & 0.031 & 0.023 & 197.6 & 164.0 & 138.6\\
correct noise & 0.079 & 0.031 & 96.2 & 84.2 & 81.0 & 0.044 & 0.021 & 196.0 & 161.3 & 137.4\\
correct prior & 0.060 & 0.033 & 97.2 & 86.3 & 82.9 & 0.036 & 0.021 & 196.5 & 162.9 & 138.0\\
correct strong prior & 0.052 & 0.035 & 97.7 & 88.1 & 84.5 & 0.036 & 0.022 & 196.9 & 164.6 & 138.2\\
unadjusted & 0.034 & 0.058 & 98.6 & 90.7 & 88.3 & 0.031 & 0.025 & 198.0 & 171.4 & 147.2\\
\addlinespace
\toprule
\multicolumn{1}{c}{ } & \multicolumn{5}{c}{Maximum sample size = 500} & \multicolumn{5}{c}{Maximum sample size = 1000} \\
\cmidrule(l{3pt}r{3pt}){2-6} \cmidrule(l{3pt}r{3pt}){7-11}
\multicolumn{3}{c}{ } & \multicolumn{3}{c}{Expected sample size} & \multicolumn{2}{c}{ } & \multicolumn{3}{c}{Expected sample size} \\
\cmidrule(l{3pt}r{3pt}){4-6} \cmidrule(l{3pt}r{3pt}){9-11}
Adjustment model & T1E & $\text{Bias}^*$ & $\gamma=1$ & $\gamma=0.72$ & $\gamma=0.60$ & T1E & $\text{Bias}^*$ & $\gamma=1$ & $\gamma=0.80$ & $\gamma=0.72$ \\
\midrule
correct & 0.028 & 0.010 & 493.7 & 404.9 & 334.9 & 0.022 & 0.008 & 992.0 & 823.2 & 681.5\\
no quad & 0.028 & 0.009 & 494.0 & 406.1 & 335.2 & 0.022 & 0.007 & 991.3 & 816.9 & 683.8\\
correct noise & 0.031 & 0.010 & 493.9 & 402.7 & 334.8 & 0.022 & 0.007 & 991.4 & 816.4 & 677.3\\
correct prior & 0.028 & 0.010 & 494.3 & 405.0 & 336.1 & 0.021 & 0.007 & 991.9 & 817.8 & 680.7\\
correct strong prior & 0.027 & 0.010 & 494.2 & 404.7 & 337.2 & 0.021 & 0.008 & 992.6 & 821.0 & 677.3\\
unadjusted & 0.026 & 0.016 & 494.6 & 426.0 & 367.0 & 0.024 & 0.010 & 990.2 & 859.6 & 734.2\\
\bottomrule
\end{tabular}}
\label{tbl:binary_table}
\end{table}

\subsection{Simulation Study Results}\label{sim_res}

\begin{figure}
    \includegraphics[width=1\textwidth]{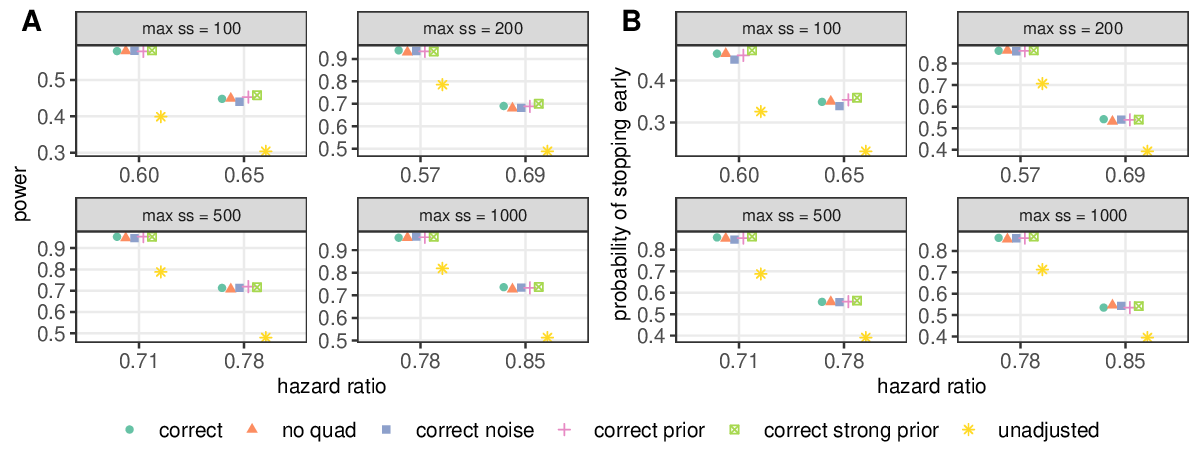}
    \caption{Time-to-event outcome. A) Power and B) probability of stopping early. Panels correspond to various maximum sample sizes (max ss). Points are jittered horizontally.}
    \centering
    \label{fig:tte_power_pse}
\end{figure}

Within each outcome type and maximum sample size, the following design operating characteristics are investigated: power, Type 1 error rate, probability of stopping early, expected sample size, posterior median bias, and RMSE. Since the sampling distribution of the test statistic $T(\mathbf{\mathcal{D}}_{n_t})$ is unknown, power and the Type 1 error rate are estimated via Monte Carlo using the 1,000 datasets. While this number is lower than that required by the FDA for adaptive simulations used in RCTs \cite{fda_iterations}, our goal here is to compare model adjustment performance, not to obtain precise estimates of operating characteristics. The probability of stopping early is estimated as the proportion of times the trial stops before performing a final analysis. In the continuous and binary outcomes, this is the proportion of times the trial stops before enrolling the maximum number of participants.  In the time-to-event outcome, this is the proportion of times the trial stops before 50 time units.  In Bayesian adaptive designs which allow for early stopping, sample size is a random variable.  Thus, expected sample size is estimated as the average of the 1,000 sample sizes at trial end.  Posterior median bias is defined as the bias resulting from using the posterior median $\hat{\gamma}$ obtained from an adjustment model as an estimator for the value of $\gamma$ used in the simulation, and is estimated through Monte Carlo using the 1,000 datasets. The Monte Carlo distribution of RMSE is displayed for each value of the marginal estimand. Here the entire posterior distribution from an adjustment model is used as the estimator for $\gamma$, so this is equivalent to the posterior expected squared error loss. For each of the 1,000 simulations, a single value of RMSE is obtained using the 3,000 posterior draws $\gamma_s$ for the value of $\gamma$ used in the simulation:
\begin{equation}
    \text{RMSE}=\sqrt{\frac{1}{3000}\sum_{s=1}^{3000}(\gamma_s - \gamma)^2}. \nonumber
\end{equation}
\noindent Results for the continuous, binary, and time-to-event endpoints are are displayed in Figures \ref{fig:cont_pow_pse}-\ref{fig:tte_power_pse} and Tables  \ref{tbl:cont_table}-\ref{tbl:tte}. For all endpoints, adjusting for variables known to be associated with the outcome increases the power of the trial and the probability of stopping the trial early as compared to the unadjusted analysis (Figures \ref{fig:cont_pow_pse}-\ref{fig:tte_power_pse}). Additionally, failing to correctly specify the functional form of a covariate (\textit{no quad}) has only a minor impact on power and the probability of stopping early. Under all scenarios, incorporating stronger prior information appears to provide little to no benefit as compared to the weakly informative priors used in the \textit{correct} models. This results from the priors being dominated by the data due to the high effective sample sizes. For the binary endpoint, this is induced by the control event risk of 0.3, which ensures that a moderately large number of events occurs throughout the trial.  For the time-to-event endpoint, this is induced by the exponential baseline hazard rate of $\lambda=0.02$, which ensures there are a large number of events within the maximum time limit of 50 time units. 

Compared to other adjustment models, adjusting for noise (\textit{correct noise}) has minimal impact under most scenarios.  However, for the smallest maximum sample size under the binary endpoint (max ss = 100), adjustment for noise slightly increases power and the probability of stopping early as compared to the \textit{correct} model. This may result from non-negligible correlation being induced between the outcome and noise variables under this setting, and so adjustment provides a further power benefit. However, this comes at the cost of a strong inflation in the Type 1 error rate as compared to the \textit{correct} model (i.e., $\text{T1E}=0.079$ versus $\text{T1E}=0.063$ in Table \ref{tbl:binary_table}). This underscores the importance of adjusting only for variables which are known to be associated with the outcome and in a pre-specified manner \cite{hauck1998should, cpmp_2004, senn2013seven, european2015guideline}. 

There is some suggestion that adjusted analyses tend to have slightly lower RMSE than unadjusted analyses under all scenarios. However, no clear pattern emerges for posterior median bias for the non-null treatment effects, where all adjustment models are comparable for practical purposes (Figures D.1-D.3 in the Online Supplementary Materials). For all scenarios except the smallest maximum sample size under the binary endpoint (max ss = 100), posterior median bias for the non-null treatment effects is negative.  This results from the estimated treatment effect being further from the null than the true treatment effect (i.e., overestimation), which is expected in trials which allow for early stopping for treatment superiority (i.e., truncated trials) \cite{mueller2007ethical, walter2019randomised, robertson2021point}. For the null treatment effect, early stopping for superiority leads to non-zero but minimal bias under all endpoints and maximum sample sizes (Tables \ref{tbl:cont_table}-\ref{tbl:tte}). For the binary endpoint, it is consistently larger in magnitude for the \textit{unadjusted} model. We note that bias under the null is negative for the continuous endpoint but positive for the binary and time-to-event endpoints. This results from the marginal difference in means being unbounded below, whereas the marginal relative risk and marginal hazard ratio are bounded below by zero.  When bias under the null is evaluated for these latter estimands on the log scale, most values become negative as in the continuous endpoint case. We elaborate further on this overestimation induced bias in Section F of the Online Supplementary Material.

Under all scenarios except the smallest maximum sample size (max ss = 100) for the binary endpoint, the Type 1 error rate is maintained below 0.05 for all adjustment models (Table \ref{tbl:cont_table}-\ref{tbl:tte}). Under the smallest maximum sample size within the binary endpoint, however, all adjustment models lead to increased Type 1 error rate as compared to the \textit{unadjusted} model.  This results from using too many covariates in the adjustment model given the low effective sample size, a phenomenon known as over-stratification \cite{kahan2014risks}.  We observe that as the maximum sample size, and thus effective sample size, increases, the inflation in the Type 1 error rate disappears.  Under the time-to-event endpoint, there is minimal inflation in the Type 1 error rate for adjusted analyses as compared to the unadjusted analysis. This may also result from over-stratification as described above. Future work should determine the optimal number of covariates to include in an adjustment model under these scenarios to avoid over-stratification. Considering the substantial power gains achieved by adjusting under the time-to-event endpoint, and that the Type 1 error rate is maintained by selecting a conservative value of the probability of superiority threshold $u$, this slight increase in the Type 1 error rate is not likely to be problematic, however. Across all non-null treatment effect scenarios, the adjusted models have lower expected sample sizes than the unadjusted model.  When combined with the probability of stopping early results, this implies that adjusted analyses are stopping more often and at earlier interim analyses for all endpoints. We note that the reduction in expected sample size is not as great for the time-to-event endpoint as compared to the other endpoints. This results from the maximum sample size being included for any interim analyses conducted past the halfway point of the trial, since all trial participants are enrolled by this point. A final simulation (included in Section E of the Online Supplementary Material) which incorporated varying degrees of prior information on the treatment effect was performed for the binary endpoint. This resulted in increased power and probability of stopping early for smaller maximum sample sizes, but at the cost of inflated type 1 error.  A more complete investigation of including informative priors on treatment effects remains as future work.

\section{Application: CCEDRRN-ADAPT}\label{covid_rct}

In this section we consider the design of a hypothetical platform trial which seeks to study the effectiveness of oral therapies against mild to moderate COVID-19 infection in individuals discharged from Canadian Emergency Departments. The trial design takes advantage of an already established network of physicians and researchers called the Canadian COVID-19 Emergency Department Rapid Response Network (CCEDRRN). We consider the first stage only, where a single oral therapy is compared to the standard of care. The binary outcome of interest is a composite endpoint of 28-day hospitalization or mortality.  Realistic values used in the trial simulation performed below are taken from a COVID-19 Emergency Department risk model, developed by the CCEDRRN researchers \cite{covid_risk_model}.  This risk model was developed using data from a high quality, population-based registry, which was also developed by the CCEDRRN researchers \cite{covid_registry, covid_outcomes_ccedrrn}.  While the binary outcome for the risk model is all cause emergency department and in-hospital mortality, this is very likely to be highly correlated with the trial's composite outcome. Thus, the simulation's results are expected to generalize to the composite endpoint as well.

The data generating mechanism for the trial simulation is shown below, where a single maximum sample size of 3,000 is chosen due to the very low marginal control event risk ($p_{ctr}=0.07$) and reflects the sample size used in CCEDRRN-ADAPT.  Letting $Y=1$ represent 28-day hospital admission or mortality, the marginal estimand of interest is the relative risk, $\gamma(\boldsymbol{\theta}) = \mu(\boldsymbol{\theta};A=1)/\mu(\boldsymbol{\theta};A=0)$ where $\mu(\boldsymbol{\theta};A=a)=E[Y|A=a;\boldsymbol{\theta}]$. We assume $\gamma(\boldsymbol{\theta})$ is intention-to-treat, that is, we compare the outcome distributions of those who are assigned to treatment versus those who are assigned to control. As above, let $\eta$ represent the linear predictor used in the data generating mechanisms.  The set $\{\boldsymbol{\beta}, \phi\}$ represents the conditional covariate and treatment effects on the log-odds scale.  As in the binary simulation case, the value of $\beta_0$ is optimized to ensure the generated datasets exhibit the correct marginal control event risk of $p_{ctr}=0.07$.  The values of $\phi$ are selected as those where the unadjusted model achieves approximately 50\% and 80\% power.  The value of $\gamma=\gamma(\boldsymbol{\theta})$ corresponding to a specific value of $\phi$ is determined through simulation, described in Appendix B of the Online Supplementary Materials. Thus, the values of $\phi$ and $\gamma$ are reported together. Let $F$ be a truncated normal distribution, parameterized by \textit{post-truncation} values of its minimum $\wedge$, maximum $\vee$, first quartile $q1$, third quartile $q3$, mean $\mu$, and standard deviation $\sigma$. An algorithm to simulate from this distribution is provided in Appendix C of the Online Supplementary Materials.  

The covariate distributions are simulated using values of summary statistics for the derivation cohort from a previously reported risk model for individuals presenting to Canadian Emergency Departments with COVID-19 symptoms \cite{covid_risk_model}.   Covariates included in the risk model and whose summary statistics were available are included in the simulation. These include age in years ($X_1$), respiratory rate upon arrival to the Emergency Department and measured in breaths/minute ($X_2$), female sex ($X_3$), chest pain ($X_4$), and arrival from police or ambulance ($X_5$). As the risk model was developed before COVID-19 vaccines were widely available, vaccination status is not included as a potential covariate.  The corresponding risk model coefficients are used as the conditional effects $\{\beta_1,...,\beta_5\}$ in the simulation. Since the summary statistics did not include covariance information, joint independence between all variables is assumed.   Due to study inclusion criteria, the range of the distribution of age is set to be $[18,90]$.  Due to biological contraints, the range of the distribution of respiratory rate is set to be $[12,40]$. Values for age and respiratory rate are simulated from the truncated normal distribution $F$ described above.

A total of 1,000 treatment-covariate datasets are generated, each containing 3,000 participants. Using these same 1,000 datasets, different outcome vectors are generated for each value of the marginal relative risk $\gamma \in \{1, 0.73, 0.63\}$, which corresponds to $\phi \in \{0, -0.42, -0.60\}$.  The outcome vectors are generated as follows:
\begin{equation}
\begin{split}
    \begin{aligned}
    Y &\sim \text{Bernoulli(logit}^{-1}(\eta)) \\
    \eta &= \beta_0 + \phi A+\beta_1X_{1} + \beta_2X_{2} + \beta_3X_{3} +\beta_4X_{4} + \beta_5X_{5} \\
    X_1 &\sim F(\wedge = 18, \vee = 90, q1=39, q3=70, \mu=54.7, \sigma=19.8) \\
    X_2 &\sim F(\wedge = 12, \vee = 40, q1=18, q3=22, \mu=21, \sigma=6.2) \\
    p_{ctr} &= 0.07  \\
    \boldsymbol{\beta} &=  (-10.76, 0.092, 0.097, -0.61, -0.80, 0.63)
    \end{aligned}
\end{split}
\hspace{10mm}
\begin{split}
    \begin{aligned}
    X_3 &\sim \text{Bernoulli}(0.478) \\
    X_4 &\sim \text{Bernoulli}(0.216) \\
    X_5 &\sim \text{Bernoulli}(0.403) \\
    \Theta &= \{\gamma \times \boldsymbol{\beta}\}.  \nonumber
    \end{aligned}
\end{split}
\end{equation}

\begin{table}
\centering
\caption{Time-to-event outcome. Type 1 error rate (T1E), bias under the null ($\text{Bias}^*$), and expected sample size at three different values of the marginal hazard ratio ($\gamma$).}
\resizebox{\linewidth}{!}{
\begin{tabular}[t]{lrrrrrrrrrr}
\toprule
\multicolumn{1}{c}{ } & \multicolumn{5}{c}{Maximum sample size = 100} & \multicolumn{5}{c}{Maximum sample size = 200} \\
\cmidrule(l{3pt}r{3pt}){2-6} \cmidrule(l{3pt}r{3pt}){7-11}
\multicolumn{3}{c}{ } & \multicolumn{3}{c}{Expected sample size} & \multicolumn{2}{c}{ } & \multicolumn{3}{c}{Expected sample size} \\
\cmidrule(l{3pt}r{3pt}){4-6} \cmidrule(l{3pt}r{3pt}){9-11}
Adjustment model & T1E & $\text{Bias}^*$ & $\gamma=1$ & $\gamma=0.65$ & $\gamma=0.60$ & T1E & $\text{Bias}^*$ & $\gamma=1$ & $\gamma=0.69$ & $\gamma=0.57$ \\
\midrule
correct & 0.027 & 0.014 & 99.7 & 98.3 & 98.0 & 0.033 & 0.000 & 199.4 & 193.1 & 189.4\\
no quad & 0.027 & 0.017 & 99.7 & 98.3 & 97.9 & 0.029 & 0.002 & 199.4 & 193.5 & 189.7\\
correct noise & 0.036 & 0.013 & 99.6 & 98.1 & 97.8 & 0.030 & -0.001 & 199.4 & 193.1 & 189.2\\
correct prior & 0.029 & 0.013 & 99.6 & 98.3 & 97.9 & 0.031 & 0.000 & 199.4 & 193.1 & 189.3\\
correct strong prior & 0.026 & 0.015 & 99.7 & 98.3 & 97.8 & 0.033 & 0.000 & 199.3 & 193.3 & 189.2\\
unadjusted & 0.029 & 0.035 & 99.7 & 98.5 & 98.2 & 0.028 & 0.005 & 199.4 & 194.4 & 191.0\\
\toprule
\multicolumn{1}{c}{ } & \multicolumn{5}{c}{Maximum sample size = 500} & \multicolumn{5}{c}{Maximum sample size = 1000} \\
\cmidrule(l{3pt}r{3pt}){2-6} \cmidrule(l{3pt}r{3pt}){7-11}
\multicolumn{3}{c}{ } & \multicolumn{3}{c}{Expected sample size} & \multicolumn{2}{c}{ } & \multicolumn{3}{c}{Expected sample size} \\
\cmidrule(l{3pt}r{3pt}){4-6} \cmidrule(l{3pt}r{3pt}){9-11}
Adjustment model & T1E & $\text{Bias}^*$ & $\gamma=1$ & $\gamma=0.78$ & $\gamma=0.71$ & T1E & $\text{Bias}^*$ & $\gamma=1$ & $\gamma=0.85$ & $\gamma=0.78$ \\
\midrule
correct & 0.026 & 0.003 & 498.4 & 477.9 & 464.8 & 0.026 & 0.001 & 996.9 & 955.4 & 912.4\\
no quad & 0.025 & 0.003 & 498.6 & 478.6 & 463.5 & 0.027 & 0.001 & 997.2 & 953.8 & 913.1\\
correct noise & 0.027 & 0.003 & 498.6 & 477.6 & 464.2 & 0.022 & 0.001 & 997.4 & 953.3 & 910.7\\
correct prior & 0.027 & 0.003 & 498.5 & 478.1 & 463.6 & 0.024 & 0.001 & 997.4 & 954.4 & 912.9\\
correct strong prior & 0.026 & 0.003 & 498.4 & 477.8 & 463.2 & 0.026 & 0.001 & 997.4 & 955.4 & 911.9\\
unadjusted & 0.022 & 0.011 & 498.6 & 483.4 & 470.6 & 0.019 & 0.003 & 997.9 & 962.8 & 929.6\\
\bottomrule
\end{tabular}}
\label{tbl:tte}
\end{table}

\noindent The trial uses the same null and alternative hypotheses as those specified under the binary endpoint simulation above and includes a probability of superiority threshold of $u=0.99$. Estimation for the marginal relative risk proceeds as previously described. Four adjustment models are considered and mirror those used in the binary simulations: \textit{correct}, \textit{correct prior}, \textit{correct strong prior}, and \textit{unadjusted}. All outcomes are assumed to be observed immediately upon participant enrollment. Interim analyses are event driven, and 75 new events are required to be observed before performing an interim analysis. This ensures a moderately high probability of stopping at an earlier interim analysis given treatment superiority.  Interim analyses continue until the trial is stopped early for superiority or until 3,000 participants are enrolled, at which time the final analysis is performed.  

Results for the CEDRRN-ADAPT design simulation study are summarized in Figure \ref{fig:covid_pow_pse} and Table \ref{tbl:covid}. Adjusting for variables known to be associated with the outcome increases the power of the trial and the probability of stopping the trial early as compared to the unadjusted analysis (Figure \ref{fig:covid_pow_pse}). As in the binary simulations above, including stronger priors on the covariate effects has minimal impact on power and the probability of stopping early as compared to the weakly informative priors used in the \textit{correct} models. While there is some suggestion that adjusted analyses tend to have slightly lower RMSE than unadjusted analyses, all adjustment models have comparable posterior median bias for the non-null treatment effects (Figure D.4 in the Online Supplementary Materials). The Type 1 error rate is maintained below 0.05, and bias under the null is minimal, for all adjustment models (Table \ref{tbl:covid}).  As in the binary simulations above, there is slight inflation in bias under the null for the \textit{unadjusted} model, however. The adjusted models have lower expected sample sizes than the unadjusted model across both non-null values of the relative risk. Again, we see the adjusted analyses are stopping more often and at earlier interim analyses as compared to the unadjusted analysis. For example, under a relative risk of 0.63, the \textit{correct} model stops at the first interim analysis 50\% of the time whereas the \textit{unadjusted} model stops at the first interim analysis only 40\% of the time.

\section{Discussion}\label{discussion}

\begin{table}
\begin{minipage}{0.4 \linewidth}
	   \centering
	   \includegraphics[width=1\textwidth]{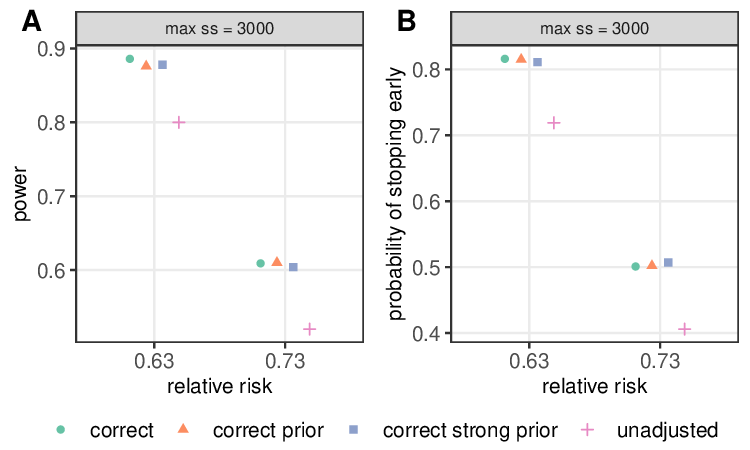}
          \captionof{figure}{COVID-19 trial with binary outcome. A) Power and B) probability of stopping early. Points are jittered horizontally.}
          \label{fig:covid_pow_pse}
	\end{minipage}\hfill
	\begin{minipage}{0.58\linewidth}
		\caption{COVID-19 trial with binary outcome. Type 1 error rate (T1E), bias under the null ($\text{Bias}^*$), and expected sample size at three different values of the marginal relative risk ($\gamma$).}
		\label{tbl:covid}
		\centering
		\resizebox{\textwidth}{!}{%
            \begin{tabular}[width=\linewidth]{@{}lrrrrr@{}}
            \toprule
            \multicolumn{3}{c}{ } & \multicolumn{3}{c}{Expected sample size} \\
            \cmidrule(l{3pt}r{3pt}){4-6}
            Adjustment model & T1E & $\text{Bias}^*$ & $\gamma=1$ & $\gamma=0.73$ & $\gamma=0.63$\\
            \midrule
            correct & 0.022 & 0.017 & 2969.7 & 2458.9 & 2047.3\\
correct prior & 0.022 & 0.017 & 2968.9 & 2455.2 & 2051.8\\
correct strong prior & 0.020 & 0.017 & 2971.5 & 2454.7 & 2056.0\\
unadjusted & 0.024 & 0.022 & 2972.7 & 2561.7 & 2220.9\\
            \bottomrule
            \end{tabular}}
        \end{minipage}
\end{table}

The impact of covariate adjustment and incorporation of prior information on covariate effects has not been previously investigated in Bayesian adaptive trials with early stopping criteria.  In this article, we assessed this impact using a variety of adjustment models and incorporated varying levels of prior information or types of model misspecification. It was shown that covariate adjustment increases power and the probability of stopping early, and decreases expected sample size over all scenarios.  Furthermore, adjusting for covariates leads to trials which stop more often and at earlier interim analyses, and can decrease RMSE as compared to unadjusted analyses. These findings are fairly robust to adjustment for noise variables, but extra caution is needed for small sample size trials (max ss = 100) with binary endpoints where noise adjustment may lead to inflated Type 1 error.  This reinforces existing best practice of only adjusting for covariates which have been pre-specified by subject matter experts \cite{hauck1998should, cpmp_2004, senn2013seven, european2015guideline}. For the scenarios explored here, which had moderate effective sample sizes, weakly informative priors on the covariate effects perform well, with stronger prior information providing little benefit. This included the CCEDRRN-ADAPT COVID-19 RCT design. Although the control event risk (0.07) of this trial is small, the maximum sample size was selected to be large enough to power the study for reasonable values of relative risk, which ensured moderately high effective sample sizes.  Including stronger prior information is expected to be helpful in trials which have small effective sample sizes (e.g., oncology trials), so future work may consider covariate adjustment within these contexts. Although we did not consider designs which also include futility stopping rules, we expect that the conclusions can be generalized.

In our simulation study, all covariates were assumed to be jointly independent. The assumption of independence may not hold in some cases and these should be investigated further. Since they carry similar information, adjusting for covariates which are moderately or highly correlated may yield smaller power increases than adjusting for approximately independent covariates. Future work might perform a simulation study to assess how different strengths of association between covariates impacts the results reported here. Another limitation of the current work is that adjustment was shown under only a single set of covariate effects within each endpoint. There is strong evidence in favor of covariate adjustment when the covariate effects are known to be strong. However, previous simulations (not shown) showed that adjustment under weaker covariate effects yielded only modest benefit. This finding was especially pronounced in the case of the non-linear endpoints. Future work might investigate a wider range of covariate effects to ascertain the magnitude that is required for covariate adjustment to be more than just modestly beneficial, though this may be context dependent.

It is important to note that the marginalization procedure used in this work is sensitive to the joint empirical distribution of the covariates with respect to which the conditional posterior samples are being marginalized.  That is, the marginalization procedure is sensitive to the participants enrolled in the trial.  While trialists work hard to obtain representative samples for use in RCTs, the participants volunteer and consent to be enrolled in the trials.  Thus they are not a random sample from the population of interest \cite{lindsey1998appropriateness}. If the sample is not as representative of the population of interest as desired, indirect standardization could be performed.  Here, conditional posterior samples would be marginalized with respect to a set of covariate patterns which more closely resembles the population of interest. Additionally, the participant covariate patterns could be augmented by \textit{pseudo-participant} covariate patterns until the sample is more representative. These could be acquired from registry data or from participants in previous trials which contained a similar population and target of interest. 

This work focused on Bayesian adaptive designs which employed simple randomization, where covariate adjustment takes place within an adaptive \textit{decision} rule. Covariate adjustment may also occur within an adaptive \textit{allocation} rule, such as in Covariate Adjusted Response Adaptive (CARA) designs \cite{rosenberger2008handling, villar2018covariate}. These designs estimate treatment arm allocation probabilities from models which include covariates.  While frequentist in nature, similar ideas can be applied within Bayesian response adaptive randomization designs. Since adaptive randomization leads to greater covariate imbalance across treatment arms, including covariate adjustment in both the adaptive decision and adaptive allocation rules may provide greater benefit than either do individually. This is an interesting direction for future work. 

The simulation study above included variants of the \textit{correct} adjustment model, which corresponded to the data generating mechanism used.  In reality, this will be unknown. Variable selection and shrinkage methods might be employed to select covariates to be used in the adjustment model. Particularly, Bayesian model averaging can be used where the decision rules would use marginal posterior samples obtained from a consensus of plausible models and might be less sensitive to any single adjustment model misspecification. However, this approach is likely to be very computationally intense. 

Another interesting direction of future research is exploration of Bayesian nonparametric models, such as Gaussian Processes, to consider more flexible functional forms that adjust for covariates.  This approach might be especially advantageous in a setting where the underlying association between the adjustment covariates and outcome is complex and hard to correctly specify, which might include a high degree of covariate non-linearity or interaction.

\section*{Data Availability}
Data sharing is not applicable to this article as no new data were created or analyzed in this study. The R scripts used for the simulations and graphics can be found on a public GitHub repository at \url{https://github.com/jjwillard/cov_adj_bact}.

\section*{Acknowledgements}

The authors would like to thank Dr. Corinne Hohl for her friendly comments and suggestions regarding the CCEDRRN-ADAPT application component of this manuscript. JW acknowledges the support of a doctoral training scholarship from the Fonds de recherche du Québec - Nature et technologies (FRQNT). SG acknowledges the support by a Discovery Grant from the Natural Sciences and Engineering Research Council of Canada (NSERC) and a Chercheurs-boursiers (Junior 1) award from the Fonds de recherche du Québec - Santé (FRQS). EEMM acknowledges support from a Discovery Grant from NSERC. EEMM is a Canada Research Chair (Tier 1) in Statistical Methods for Precision Medicine and acknowledges the support of a chercheur de mérite career award from the FRQS. All simulations were performed on the Narval high performance computing cluster, located at the  École de technologie supérieure in Montréal, Canada. JW thanks the research staff at Calcul Québec and the Digital Research Alliance of Canada (formerly Compute Canada) for their technical assistance.

\bibliographystyle{apalike}  
\bibliography{references.bib}  

\begin{thebibliography}{}

\bibitem[Austin et~al., 2010]{austin2010substantial}
Austin, P.~C., Manca, A., Zwarenstein, M., Juurlink, D.~N., and Stanbrook,
  M.~B. (2010).
\newblock {A Substantial and Confusing Variation Exists in Handling of Baseline
  Covariates in Randomized Controlled Trials: a Review of Trials Published in
  Leading Medical Journals}.
\newblock {\em Journal of Clinical Epidemiology}, 63(2):142--153.

\bibitem[Benkeser et~al., 2021]{benkeser2021improving}
Benkeser, D., D{\'\i}az, I., Luedtke, A., Segal, J., Scharfstein, D., and
  Rosenblum, M. (2021).
\newblock {Improving Precision and Power in Randomized Trials for COVID-19
  Treatments Using Covariate Adjustment, for Binary, Ordinal, and Time-to-event
  Outcomes}.
\newblock {\em Biometrics}, 77(4):1467--1481.

\bibitem[Berry et~al., 2010]{berry2010bayesian}
Berry, S.~M., Carlin, B.~P., Lee, J.~J., and Muller, P. (2010).
\newblock {\em {Bayesian Adaptive Methods for Clinical Trials}}.
\newblock CRC press.

\bibitem[Brilleman et~al., 2020]{brilleman2020bayesian}
Brilleman, S.~L., Elci, E.~M., Novik, J.~B., and Wolfe, R. (2020).
\newblock {Bayesian Survival Analysis Using the rstanarm R Package}.
\newblock {\em arXiv preprint arXiv:2002.09633}.

\bibitem[Ciolino et~al., 2019]{ciolino2019ideal}
Ciolino, J.~D., Palac, H.~L., Yang, A., Vaca, M., and Belli, H.~M. (2019).
\newblock {Ideal vs. Real: a Systematic Review on Handling Covariates in
  Randomized Controlled Trials}.
\newblock {\em BMC Medical Research Methodology}, 19(1):1--11.

\bibitem[{Committee for Proprietary Medicinal Products}\ignorespaces,
  2004]{cpmp_2004}
{Committee for Proprietary Medicinal Products}\ignorespaces (2004).
\newblock {Points to Consider on Adjustment for Baseline Covariates}.
\newblock {\em Statistics in Medicine}, 23(5):701--709.

\bibitem[Daniel et~al., 2021]{daniel2021making}
Daniel, R., Zhang, J., and Farewell, D. (2021).
\newblock {Making Apples from Oranges: Comparing Noncollapsible Effect
  Estimators and their Standard Errors after Adjustment for Different Covariate
  Sets}.
\newblock {\em Biometrical Journal}, 63(3):528--557.

\bibitem[Daniels et~al., 2023]{daniels2023bayesian}
Daniels, M.~J., Linero, A., and Roy, J. (2023).
\newblock {\em {Bayesian Nonparametrics for Causal Inference and Missing
  Data}}.
\newblock CRC Press, Boca Raton, FL.

\bibitem[{European Medicines Agency}, 2015]{european2015guideline}
{European Medicines Agency} (2015).
\newblock {Guideline on Adjustment for Baseline Covariates in Clinical Trials}.
\newblock
  https://www.ema.europa.eu/en/documents/scientific-guideline/guideline-adjustment-baseline-covariates-clinical-trials\_en.pdf
  (accessed 12 May 2022).

\bibitem[Freeman~Jr and Holford, 1980]{freeman1980summary}
Freeman~Jr, D.~H. and Holford, T.~R. (1980).
\newblock {Summary Rates}.
\newblock {\em Biometrics}, 36(2):195--205.

\bibitem[Gabry and Goodrich, 2022]{rstanarm_priors}
Gabry, J. and Goodrich, B. (2022).
\newblock {Prior Distributions for rstanarm Models}.
\newblock
  https://cran.r-project.org/web/packages/rstanarm/vignettes/priors.html
  (accessed 21 April 2022).

\bibitem[Gelman and Rubin, 1992]{rhat_gelman_rubin}
Gelman, A. and Rubin, D.~B. (1992).
\newblock {Inference from Iterative Simulation Using Multiple Sequences}.
\newblock {\em Statistical Science}, 7(4):457--472.

\bibitem[Giovagnoli, 2021]{giovagnoli2021bayesian}
Giovagnoli, A. (2021).
\newblock {The Bayesian Design of Adaptive Clinical Trials}.
\newblock {\em International Journal of Environmental Research and Public
  Health}, 18(2):530.

\bibitem[Goodrich et~al., 2020]{rstanarm}
Goodrich, B., Gabry, J., Ali, I., and Brilleman, S. (2020).
\newblock {rstanarm: {Bayesian} Applied Regression Modeling via {Stan}}.
\newblock https://mc-stan.org/rstanarm (version 2.21.2).

\bibitem[Harrell and Slaughter, 2021]{harrell2021bbr}
Harrell, F. and Slaughter, J. (2021).
\newblock {Biostatistics for Biomedical Research}.
\newblock \text{http://hbiostat.org/doc/bbr.pdf (accessed 03 Feb 2022)}.

\bibitem[Hauck et~al., 1998]{hauck1998should}
Hauck, W.~W., Anderson, S., and Marcus, S.~M. (1998).
\newblock {Should We Adjust for Covariates in Nonlinear Regression Analyses of
  Randomized Trials?}
\newblock {\em Controlled Clinical Trials}, 19(3):249--256.

\bibitem[Hernández et~al., 2006]{hernandez_2006_tte_adj}
Hernández, A.~V., Eijkemans, M.~J., and Steyerberg, E.~W. (2006).
\newblock {Randomized Controlled Trials With Time-to-Event Outcomes: How Much
  Does Prespecified Covariate Adjustment Increase Power?}
\newblock {\em Annals of Epidemiology}, 16(1):41--48.

\bibitem[Hernández et~al., 2004]{hernandez_2004_cov_adj}
Hernández, A.~V., Steyerberg, E.~W., and Habbema, J.~F. (2004).
\newblock {Covariate Adjustment in Randomized Controlled Trials with
  Dichotomous Outcomes Increases Statistical Power and Reduces Sample Size
  Requirements}.
\newblock {\em Journal of Clinical Epidemiology}, 57(5):454--460.

\bibitem[Hohl et~al., 2022a]{covid_risk_model}
Hohl, C.~M., Rosychuk, R.~J., Archambault, P.~M., O’Sullivan, F., Leeies, M.,
  Mercier, {\'E}., Clark, G., Innes, G.~D., Brooks, S.~C., Hayward, J., et~al.
  (2022a).
\newblock {The CCEDRRN COVID-19 Mortality Score to Predict Death Among
  Nonpalliative Patients with COVID-19 Presenting to Emergency Departments: a
  Derivation and Validation Study}.
\newblock {\em Canadian Medical Association Open Access Journal},
  10(1):E90--E99.

\bibitem[Hohl et~al., 2022b]{covid_outcomes_ccedrrn}
Hohl, C.~M., Rosychuk, R.~J., Hau, J.~P., Hayward, J., Landes, M., Yan, J.~W.,
  Ting, D.~K., Welsford, M., Archambault, P.~M., Mercier, E., Chandra, K.,
  Davis, P., Vaillancourt, S., Leeies, M., Small, S., and Morrison, L.~J.
  (2022b).
\newblock {Treatments, Resource Utilization, and Outcomes of COVID-19 Patients
  Presenting to Emergency Departments Across Pandemic Waves: an Observational
  Study by the Canadian COVID-19 Emergency Department Rapid Response Network
  (CCEDRRN)}.
\newblock {\em Canadian Journal of Emergency Medicine}, 24:397–407.

\bibitem[Hohl et~al., 2021]{covid_registry}
Hohl, C.~M., Rosychuk, R.~J., McRae, A.~D., Brooks, S.~C., Archambault, P.,
  Fok, P.~T., Davis, P., Jelic, T., Turner, J.~P., Rowe, B.~H., Mercier,
  {\'E}., Cheng, I., Taylor, J., Daoust, R., Ohle, R., Andolfatto, G., Atzema,
  C., Hayward, J., Khangura, J.~K., Landes, M., Lang, E., Martin, I., Mohindra,
  R., Ting, D.~K., Vaillancourt, S., Welsford, M., Brar, B., Dahn, T., Wiemer,
  H., Yadav, K., Yan, J.~W., Stachura, M., McGavin, C., Perry, J.~J., and
  Morrison, L.~J. (2021).
\newblock {Development of the Canadian COVID-19 Emergency Department Rapid
  Response Network Population-based Registry: a Methodology Study}.
\newblock {\em Canadian Medical Association Open Access Journal},
  9(1):E261--E270.

\bibitem[Kahan et~al., 2014]{kahan2014risks}
Kahan, B.~C., Jairath, V., Dor{\'e}, C.~J., and Morris, T.~P. (2014).
\newblock {The Risks and Rewards of Covariate Adjustment in Randomized Trials:
  an Assessment of 12 Outcomes from 8 Studies}.
\newblock {\em Trials}, 15(1):1--7.

\bibitem[Kalton, 1968]{kalton1968standardization}
Kalton, G. (1968).
\newblock {Standardization: A Technique to Control for Extraneous Variables}.
\newblock {\em Journal of the Royal Statistical Society: Series C (Applied
  Statistics)}, 17(2):118--136.

\bibitem[Keil et~al., 2018]{keil2018bayesian}
Keil, A.~P., Daza, E.~J., Engel, S.~M., Buckley, J.~P., and Edwards, J.~K.
  (2018).
\newblock {A Bayesian Approach to the G-formula}.
\newblock {\em Statistical Methods in Medical Research}, 27(10):3183--3204.

\bibitem[Lane and Nelder, 1982]{lane1982analysis}
Lane, P.~W. and Nelder, J.~A. (1982).
\newblock {Analysis of Covariance and Standardization as Instances of
  Prediction}.
\newblock {\em Biometrics}, 38(3):613--621.

\bibitem[Lee et~al., 2022]{lee2022benefits}
Lee, K.~M., Robertson, D.~S., Jaki, T., and Emsley, R. (2022).
\newblock {The Benefits of Covariate Adjustment for Adaptive Multi-arm
  Designs}.
\newblock {\em Statistical Methods in Medical Research}, 31(11):2104--2121.

\bibitem[Lewis, 1999]{ich1999harmonised}
Lewis, J.~A. (1999).
\newblock {Statistical Principles for Clinical Trials (ICH E9): An Introductory
  Note on an International Guideline}.
\newblock {\em Statistics in Medicine}, 18(15):1903--1942.

\bibitem[Lindsey and Lambert, 1998]{lindsey1998appropriateness}
Lindsey, J.~K. and Lambert, P. (1998).
\newblock {On the Appropriateness of Marginal Models for Repeated Measurements
  in Clinical Trials}.
\newblock {\em Statistics in Medicine}, 17(4):447--469.

\bibitem[Linero and Antonelli, 2023]{linero2023and}
Linero, A.~R. and Antonelli, J.~L. (2023).
\newblock {The How and Why of Bayesian Nonparametric Causal Inference}.
\newblock {\em Wiley Interdisciplinary Reviews: Computational Statistics},
  15(1):e1583.

\bibitem[Mueller et~al., 2007]{mueller2007ethical}
Mueller, P.~S., Montori, V.~M., Bassler, D., Koenig, B.~A., and Guyatt, G.~H.
  (2007).
\newblock {Ethical Issues in Stopping Randomized Trials Early Because of
  Apparent Benefit}.
\newblock {\em Annals of Internal Medicine}, 146(12):878--881.

\bibitem[Oganisian and Roy, 2021]{oganisian2021practical}
Oganisian, A. and Roy, J.~A. (2021).
\newblock {A Practical Introduction to Bayesian Estimation of Causal Effects:
  Parametric and Nonparametric Approaches}.
\newblock {\em Statistics in Medicine}, 40(2):518--551.

\bibitem[{R Core Team}, 2020]{r_proj}
{R Core Team} (2020).
\newblock {R: A Language and Environment for Statistical Computing}.
\newblock https://www.R-project.org/ (version 4.2.1).

\bibitem[Remiro-Az{\'o}car et~al., 2020]{remiro2020marginalization}
Remiro-Az{\'o}car, A., Heath, A., and Baio, G. (2020).
\newblock {Marginalization of Regression-Adjusted Treatment Effects in Indirect
  Comparisons with Limited Patient-Level Data}.
\newblock {\em arXiv preprint arXiv:2008.05951}.

\bibitem[Robertson et~al., 2022]{robertson2021point}
Robertson, D.~S., Choodari-Oskooei, B., Dimairo, M., Flight, L., Pallmann, P.,
  and Jaki, T. (2022).
\newblock {Point Estimation for Adaptive Trial Designs I: A Methodological
  Review}.
\newblock {\em Statistics in Medicine}, Advance Online Publication.

\bibitem[Rosenberger and Sverdlov, 2008]{rosenberger2008handling}
Rosenberger, W.~F. and Sverdlov, O. (2008).
\newblock {Handling Covariates in the Design of Clinical Trials}.
\newblock {\em Statistical Science}, 23(3):404--419.

\bibitem[Rubin, 1981]{rubin1981bayesian}
Rubin, D.~B. (1981).
\newblock {The Bayesian Bootstrap}.
\newblock {\em The Annals of Statistics}, 9(1):130--134.

\bibitem[Saarela et~al., 2015]{saarela2015predictive}
Saarela, O., Arjas, E., Stephens, D.~A., and Moodie, E. (2015).
\newblock {Predictive Bayesian Inference and Dynamic Treatment Regimes}.
\newblock {\em Biometrical Journal}, 57(6):941--958.

\bibitem[Senn, 2013]{senn2013seven}
Senn, S. (2013).
\newblock {Seven Myths of Randomisation in Clinical Trials}.
\newblock {\em Statistics in Medicine}, 32(9):1439--1450.

\bibitem[Stephens et~al., 2022]{stephens2022causal}
Stephens, D., Nobre, W., Moodie, E., and Schmidt, A. (2022).
\newblock {Causal Inference Under Mis-specification: Adjustment Based on the
  Propensity Score}.
\newblock {\em Bayesian Analysis}, 1(1):1--24.

\bibitem[Stitelman et~al., 2011]{stitelman2011targeted}
Stitelman, O.~M., Wester, C.~W., De~Gruttola, V., and {van der Laan}, M.~J.
  (2011).
\newblock {Targeted Maximum Likelihood Estimation of Effect Modification
  Parameters in Survival Analysis}.
\newblock {\em The International Journal of Biostatistics}, 7(1).

\bibitem[{US Food and Drug Administration}, 2019]{fda_iterations}
{US Food and Drug Administration} (2019).
\newblock {Adaptive Designs for Clinical Trials of Drugs and Biologics:
  Guidance for Industry}.
\newblock https://www.fda.gov/media/78495/download (accessed 01 Dec 2022).

\bibitem[{US Food and Drug Administration}, 2021]{food148910adjusting}
{US Food and Drug Administration} (2021).
\newblock {Adjusting for Covariates in Randomized Clinical Trials for Drugs and
  Biologics with Continuous Outcomes: Guidance for Industry}.
\newblock https://www.fda.gov/media/148910/download (accessed 12 May 2022).

\bibitem[Van~Lancker et~al., 2022]{van2022combining}
Van~Lancker, K., Betz, J., and Rosenblum, M. (2022).
\newblock {Combining Covariate Adjustment with Group Sequential and Information
  Adaptive Designs to Improve Randomized Trial Efficiency}.
\newblock {\em arXiv preprint arXiv:2201.12921}.

\bibitem[Villar and Rosenberger, 2018]{villar2018covariate}
Villar, S. and Rosenberger, W. (2018).
\newblock {Covariate-Adjusted Response-Adaptive Randomization for Multi-Arm
  Clinical Trials Using a Modified Forward Looking Gittins Index Rule}.
\newblock {\em Biometrics}, 74(1):49--57.

\bibitem[Walter et~al., 2019]{walter2019randomised}
Walter, S., Guyatt, G., Bassler, D., Briel, M., Ramsay, T., and Han, H. (2019).
\newblock {Randomised Trials with Provision for Early Stopping for Benefit (or
  Harm): the Impact on the Estimated Treatment Effect}.
\newblock {\em Statistics in Medicine}, 38(14):2524--2543.

\bibitem[Wang and Yan, 2021]{splines2}
Wang, W. and Yan, J. (2021).
\newblock {Shape-Restricted Regression Splines with {R} Package {splines2}}.
\newblock {\em Journal of Data Science}, 19(3):498--517.

\end{thebibliography}

\newpage
\begin{singlespace}
\appendixpage
\setcounter{page}{1}
\setcounter{section}{1}
\begin{appendices}
\textbf{James Willard, Shirin Golchi, and Erica EM Moodie}
\section{Marginalization Procedures}\label{marginalization_procedures}
\counterwithin{figure}{section}

\subsection{Continuous Outcome: Difference in Means}\label{alg:marg_continuous}

Let $Y$ be a Normally distributed outcome with the target of inference being the marginal difference in means:
\begin{equation}
    \gamma(\boldsymbol{\theta}) := \mu(\boldsymbol{\theta};A=1) - \mu(\boldsymbol{\theta};A=0) \nonumber
\end{equation}
\noindent where $\mu(\boldsymbol{\theta};A=a) = E[Y \mid A=a;\boldsymbol{\theta}]$. 
Under the assumption of at least one treatment-covariate interaction (i.e., $\mathbf{Z} \ne \emptyset$; treatment effect heterogeneity), the difference in means is non-collapsible. Estimation proceeds assuming independent outcomes and the following model:    
\begin{equation}
\begin{split}
    \begin{aligned}
        &p(Y_i \mid A_i,\mathbf{X}_i,\boldsymbol{\theta}) = \text{Normal}(\mu(\boldsymbol{\theta};A_i,\mathbf{X}_i),\sigma^2) \\
        &\mu(\boldsymbol{\theta};A_i,\mathbf{X}_i) = \beta_0 + \phi A_i + \mathbf{X_i}\boldsymbol{\beta} + (A_i \cdot \mathbf{Z_i})\boldsymbol{\omega} \\
        &\boldsymbol{\theta} = \{\beta_0, \phi, \boldsymbol{\beta}, \boldsymbol{\omega}, \sigma^2\} 
    \end{aligned}
\end{split}
\hspace{10mm}
\begin{split}
    \begin{aligned}
        &\mathcal{L} = \prod_{i=1}^{n_t} p(Y_i \mid A_i,\mathbf{X}_i,\boldsymbol{\theta}) \\
        &\pi(\boldsymbol{\theta} \mid \mathcal{D}_{n_t}) \propto \prod_{i=1}^{n_t} p(Y_i \mid A_i,\mathbf{X}_i,\boldsymbol{\theta}) p(\boldsymbol{\theta}).
        \nonumber
    \end{aligned}
    \end{split}
\end{equation}
Samples from the posterior distribution of $\gamma(\boldsymbol{\theta})$ are obtained from adjusted analyses by marginalizing $s=1,...,S$ samples of the conditional $\mu(\boldsymbol{\theta};A,\mathbf{X})$ before forming the contrast:
\begin{equation}
\begin{aligned}
\gamma(\boldsymbol{\theta}_s) &= \mu(\boldsymbol{\theta}_s;A=1) - \mu(\boldsymbol{\theta}_s;A=0) \\
&= \int_{\mathbf{X}} \mu(\boldsymbol{\theta}_s;A=1,\mathbf{X})p(\mathbf{X})d\mathbf{X} - \int_{\mathbf{X}} \mu(\boldsymbol{\theta}_s;A=0,\mathbf{X})p(\mathbf{X})d\mathbf{X} \\
&= \int_{\mathbf{X}}(\beta_{0,s} + \phi_s + \mathbf{X}\boldsymbol{\beta}_s +  \mathbf{Z}\boldsymbol{\omega}_s)p(\mathbf{X})d\mathbf{X} - \int_{\mathbf{X}}(\beta_{0,s} + \mathbf{X}\boldsymbol{\beta}_s)p(\mathbf{X})d\mathbf{X}. \nonumber
\end{aligned}   
\end{equation}
\noindent The integrals are approximated using the Bayesian bootstrap procedure described in Section \ref{marg_procedures} of the manuscript. After fitting the linear model, $s=1,...,S$ samples are obtained from the joint posterior distribution of the model parameters, $\pi(\boldsymbol{\theta} \mid \mathcal{D}_{n_t})$.  Let $\boldsymbol{\theta}_s$ represent the $s^{th}$ draw from this joint posterior distribution.  For every row $i=1,...,{n_t}$ in the sample data, a value of $A_i=1$ is assigned.  Then for each $\boldsymbol{\theta}_s$ the following procedure is performed. The $n_t$ values of the linear predictor $\mu(\boldsymbol{\theta}_s;A_i=1,\mathbf{X}_i=\mathbf{x}_i)$ are calculated. A vector $\mathbf{w}_s=(w_{1,s},...,w_{n_t,s})$ is drawn from a Dirichlet($\mathbf{1}_{n_t}$) distribution.  Using $\mathbf{w}_s$, the $n_t$ values are then averaged, $\sum_{i=1}^{n_t}w_{i,s}\mu(\boldsymbol{\theta}_s;A_i=1,\mathbf{X}_i=\mathbf{x}_i)$, which marginalizes them with respect to the observed $\mathbf{X}=\mathbf{x}$, yielding a single sample $\mu(\boldsymbol{\theta}_s;A=1)$.  This occurs for all $\boldsymbol{\theta}_s$ to yield $S$ samples from the posterior distribution of $\mu(\boldsymbol{\theta};A=1)$.  This entire process is then repeated for $A_i=0$, to yield $S$ samples from the posterior distribution of $\mu(\boldsymbol{\theta};A=0)$.  These posterior samples are then subtracted to yield samples from the posterior distribution of $\gamma(\boldsymbol{\theta})$. A brief summary outline is included below.
\begin{enumerate}
    \item Fit the linear regression model with identity link.
    \item Obtain $s=1,...,S$ samples from the joint posterior distribution of the model parameters, $\pi(\boldsymbol{\theta} \mid \mathcal{D}_{n_t})$.
    \item \label{lab1} Create one copy of the sample data where $A_i=1$ for all $i=1,...,{n_t}$.
    \item For each $\boldsymbol{\theta}_s$, perform the following: \label{lab2}
    \begin{enumerate}
        \item  For each $i=1,...,{n_t}$, calculate ${\mu(\boldsymbol{\theta}_s;A_i=1,\mathbf{X}_i=\mathbf{x}_i)}$. 
        \item Sample $\mathbf{w}_s=(w_{1,s},...,w_{n_t,s})$ from a Dirichlet($\mathbf{1}_{n_t}$) distribution.
        \item \label{lab3} Average these $n_t$ values to marginalize with respect to the observed $\mathbf{X}=\mathbf{x}$: \\
        $\mu(\boldsymbol{\theta}_s;A_i=1)=\sum_{i=1}^{n_t}w_{i,s}\mu(\boldsymbol{\theta}_s;A_i=1,\mathbf{X}_i=\mathbf{X}_i)$.
    \end{enumerate}
    \item The $s=1,...,S$ values of ${\mu(\boldsymbol{\theta}_s;A=1)}$ are samples from the posterior of $\mu(\boldsymbol{\theta};A=1)$.
    \item Repeat steps \ref{lab1}-\ref{lab2} for $A_i=0$ to yield $S$ samples from the posterior of $\mu(\boldsymbol{\theta};A=0)$.
    \item Subtract to obtain $S$ samples from the posterior of $\gamma(\boldsymbol{\theta})$.
\end{enumerate}

\subsection{Binary Outcome: Relative Risk}\label{alg:marg_binary}

\noindent Letting $Y$ be distributed as a Bernoulli random variable, where $Y=1$ indicates an event occurs and $Y=0$ indicates no event occurs, a marginal estimand of interest is the relative risk:
\begin{equation}
\gamma(\boldsymbol{\theta}) := \mu(\boldsymbol{\theta};A=1)/\mu(\boldsymbol{\theta};A=0) \nonumber
\end{equation}
\noindent where $\mu(\boldsymbol{\theta};A=a)=E[Y\mid A=a;\boldsymbol{\theta}]$. Estimation proceeds assuming independent outcomes and the following model:  
\begin{equation}
\begin{split}
\begin{aligned}
        &p(Y_i \mid A_i,\mathbf{X}_i,\boldsymbol{\theta}) = \text{Bernoulli}(\mu(\boldsymbol{\theta};A_i,\mathbf{X}_i)) \\
        &\mu(\boldsymbol{\theta};A_i,\mathbf{X}_i) =  \text{logit}^{-1}(\beta_0 + \phi A_i + \mathbf{X}_i\boldsymbol{\beta} + (A_i \cdot \mathbf{Z}_i)\boldsymbol{\omega}) \\
        &\boldsymbol{\theta}=\{\beta_0, \phi, \boldsymbol{\beta}, \boldsymbol{\omega}\} 
\end{aligned}
\end{split}
\hspace{10mm}
\begin{split}
    \begin{aligned}
    &\mathcal{L} = \prod_{i=1}^{n_t} p(Y_i \mid A_i,\mathbf{X}_i,\boldsymbol{\theta}) \\
    &\pi(\boldsymbol{\theta} \mid \mathcal{D}_{n_t}) \propto \prod_{i=1}^{n_t} p(Y_i \mid A_i,\mathbf{X}_i,\boldsymbol{\theta})p(\boldsymbol{\theta}).
    \nonumber
    \end{aligned}
\end{split}
\end{equation}
To obtain posterior samples of the marginal estimand from adjusted analyses, $s=1,...,S$ posterior samples from the inverse logit link function applied to the linear predictors under treatment and no treatment are marginalized and then divided:
\begin{equation}
\begin{aligned}
     \gamma(\boldsymbol{\theta}_s) &= \frac{\mu(\boldsymbol{\theta}_s;A=1)}{\mu(\boldsymbol{\theta}_s;A=0)} \\ 
     &= \frac{\int_{\mathbf{X}}\mu(\boldsymbol{\theta}_s;A=1,\mathbf{X})p(\mathbf{X})d(\mathbf{X})}{\int_{\mathbf{X}}\mu(\boldsymbol{\theta}_s;A=0,\mathbf{X})p(\mathbf{X})d(\mathbf{X})} \\
     &= \frac{\int_{\mathbf{X}}\text{logit}^{-1}\{\beta_{0,s} + \phi_s + \mathbf{X}\boldsymbol{\beta}_s + \mathbf{Z}\boldsymbol{\omega}_s\}p(\mathbf{X})d(\mathbf{X})}{\int_{\mathbf{X}}\text{logit}^{-1}\{\beta_{0,s} + \mathbf{X}\boldsymbol{\beta}_s\}p(\mathbf{X})d(\mathbf{X})} \nonumber
\end{aligned}
\end{equation}
\noindent The integrals are approximated using the Bayesian bootstrap procedure described in Section \ref{marg_procedures} of the manuscript. After fitting the generalized linear model with logit link function corresponding to $\eta$, $s=1,...,S$ samples are obtained from the joint posterior distribution of the model parameters, $\pi(\boldsymbol{\theta} \mid \mathcal{D}_{n_t})$. Let $\boldsymbol{\theta}_s$ represent the $s^{th}$ draw from this joint posterior distribution.  For every row $i = 1,...,{n_t}$ in the sample data, a value of $A_i=1$ is assigned.  Then for each $\boldsymbol{\theta}_s$ the following procedure is performed. The $n_t$ values of the indexed linear predictors are calculated and transformed by the inverse logit to yield samples from $\mu(\boldsymbol{\theta}_s;A_i=1,\mathbf{X}_i=\mathbf{x}_i)=\text{logit}^{-1}(\beta_{0,s} + \phi_s A_i + \mathbf{X}_i\boldsymbol{\beta}_s + (A_i \cdot \mathbf{Z}_i)\boldsymbol{\omega}_s)$. A vector $\mathbf{w}_s=(w_{1,s},...,w_{n_t,s})$ is drawn from a Dirichlet($\mathbf{1}_{n_t}$) distribution.  Using $\mathbf{w}_s$, the $n_t$ values are then averaged, ${\sum_{i=1}^{n_t}w_{i,s}\mu(\boldsymbol{\theta}_s;A_i=1,\mathbf{X}_i=\mathbf{x}_i)}$, which marginalizes them with respect to the observed $\mathbf{X}=\mathbf{x}$, yielding a single sample $\mu(\boldsymbol{\theta}_s;A=1)$ from the posterior distribution of $\mu(\boldsymbol{\theta};A=1)$.  This occurs for all $\boldsymbol{\theta}_s$ to yield $S$ samples from the posterior distribution of $\mu(\boldsymbol{\theta};A=1)$.  This entire process is then repeated for $A_i=0$, to yield $S$ samples from the posterior distribution of $\mu(\boldsymbol{\theta};A=0)$.  These are then divided to yield $S$ samples from the posterior distribution of $\gamma(\boldsymbol{\theta})$. A brief summary outline is included below.
\begin{enumerate}
    \item Fit the logistic regression model.
    \item Obtain $s=1,...,S$ samples from the joint posterior distribution of the model parameters, $\pi(\boldsymbol{\theta} \mid \mathcal{D}_{n_t})$.
    \item \label{bin_lab1} Create one copy of the sample data where $A_i=1$ for all $i=1,...,n_t$.
    \item For each $\boldsymbol{\theta}_s$, perform the following: \label{bin_lab2}
    \begin{enumerate}
        \item  For each $i=1,...,n_t$, calculate $\mu(\boldsymbol{\theta}_s;A_i=1,\mathbf{X}_i=\mathbf{x}_i)=\text{logit}^{-1}(\beta_{0,s} + \phi_s A_i + \mathbf{X}_i\boldsymbol{\beta}_s + (A_i \cdot \mathbf{Z}_i)\boldsymbol{\omega}_s)$.
        \item Sample $\mathbf{w}_s=(w_{1,s},...,w_{n_t,s})$ from a Dirichlet($\mathbf{1}_{n_t}$) distribution.
        \item \label{bin_lab3} Average these $n_t$ values to marginalize with respect to the observed $\mathbf{X}=\mathbf{x}$: \\ $\mu(\boldsymbol{\theta}_s;A=1)=\sum_{i=1}^{n_t}w_{i,s}\mu(\boldsymbol{\theta}_s;A_i=1,\mathbf{X}_i=\mathbf{x}_i)$.
    \end{enumerate}
    \item The $s=1,...,S$ values of ${\mu(\boldsymbol{\theta}_s;A=1)}$ are samples from the posterior distribution of $\mu(\boldsymbol{\theta};A=1)$.
    \item Repeat steps \ref{bin_lab1}-\ref{bin_lab2} for $A_i=0$ to yield $S$ samples from the posterior distribution of $\mu(\boldsymbol{\theta};A=0)$.
    \item Divide ${\mu(\boldsymbol{\theta}_s;A=1)/{\mu(\boldsymbol{\theta}_s;A=0)}}$ for each $s$ to obtain $S$ samples from the posterior distribution of $\gamma(\boldsymbol{\theta})$.
\end{enumerate}

\subsection{Time-to-event Outcome: Marginalization of Conditional Hazard Ratio Estimates}\label{alg:marg_tte}

Let $Y=\{T,\delta\}$ be defined as in the section for hazard ratios in the manuscript, where the target of inference is the marginal hazard ratio:
\begin{equation}
\begin{aligned}
   \gamma(\boldsymbol{\theta}) &= h(t \mid A=1)/ h(t \mid A=0) \\
   &=\log\{\mu(\boldsymbol{\theta};A=1)\}/\log\{\mu(\boldsymbol{\theta};A=0)\} \nonumber 
\end{aligned}
\end{equation}
\noindent where $\mu(\boldsymbol{\theta};A=a)=S(t\mid A=a;\boldsymbol{\theta})$. Estimation proceeds assuming independent outcomes, no competing risks, and the following model:
\begin{equation}
\begin{split}
\begin{aligned}
    &h_i(t \mid A_i, \mathbf{X}_i)=h_0(t)\exp(\eta_i) \\
    &\eta_i = \phi A_i + \mathbf{X}_i \boldsymbol{\beta} + (A_i \cdot \mathbf{Z_i})\boldsymbol{\omega} \\
    &S_i(t \mid A_i, \mathbf{X}_i) =\exp\left(-I(t;\boldsymbol{\psi}, \boldsymbol{k},\delta)\exp(\eta_i)\right) \\
    &\boldsymbol{\theta}=\{\boldsymbol{\psi}, \phi, \boldsymbol{\beta}, \boldsymbol{\omega}\} 
\end{aligned}
\end{split}
\hspace{10mm}
\begin{split}
    \begin{aligned}
    &p(Y_i \mid A_i, \mathbf{X}_i, \boldsymbol{\theta}) = S_i(T_i \mid A_i, \mathbf{X}_i)^{1-\delta_i}h_i(T_i \mid A_i, \mathbf{X}_i)^{\delta_i} \\
    &\mathcal{L} = \prod_{i=1}^{n_t} p(Y_i \mid A_i, \mathbf{X}_i, \boldsymbol{\theta}) \\
    &\pi(\boldsymbol{\theta} \mid \mathcal{D}_{n_t}) \propto \prod_{i=1}^{n_t} p(Y_i \mid A_i, \mathbf{X}_i, \boldsymbol{\theta})p(\boldsymbol{\theta}). 
    \nonumber
    \end{aligned}
\end{split}
\end{equation}
As the hazard ratio is non-collapsible, $s=1,...,S$ marginal posterior samples from adjusted analyses are obtained through marginalization of the log transformed survival probabilities:
\begin{equation}
    \begin{aligned}
    \gamma(\boldsymbol{\theta}_s) &= \frac{h_s(t \mid A=1)}{h_s(t \mid A=0)} \\
    &= \frac{\log\{\mu(\boldsymbol{\theta}_s;A=1)\}}{\log\{\mu(\boldsymbol{\theta}_s;A=0)\}} \\
    &= \frac{\log\{\int_{\mathbf{X}}\mu(\boldsymbol{\theta}_s;A=1,\mathbf{X})p(\mathbf{X})d\mathbf{X}\}}{\log\{\int_{\mathbf{X}}\mu(\boldsymbol{\theta}_s;A=0,\mathbf{X})p(\mathbf{X})d\mathbf{X}\}} \\
    &= \frac{\log\{\int_{\mathbf{X}}\exp\left[-I(t;\boldsymbol{\psi}_s, \boldsymbol{k},\delta)\exp(\phi_s + \mathbf{X} \boldsymbol{\beta}_s + \mathbf{Z}\boldsymbol{\omega}_s)\right]p(\mathbf{X})d\mathbf{X}\}}{\log\{\int_{\mathbf{X}}\exp\left[-I(t;\boldsymbol{\psi}_s, \boldsymbol{k},\delta)\exp(\mathbf{X} \boldsymbol{\beta}_s)\right]p(\mathbf{X})d\mathbf{X}\}}. \nonumber
    \end{aligned}
\end{equation}

\noindent Dividing log-transformed survival probabilities can be numerically unstable and is undefined for all $t$ such that $S(t|A=a) \in \{0,1\}$. Thus we use a more numerically stable identity \cite{stitelman2011targeted, remiro2020marginalization}:
\begin{equation}
    \begin{aligned}
\log(\gamma(\boldsymbol{\theta}_s))&=\log \left( \frac{h_s(t|A=1)}{h_s(t|A=0)} \right)=\log \left( \frac{-\log[\mu(\boldsymbol{\theta}_s;A=1)]}{-\log[\mu(\boldsymbol{\theta}_s;A=0)]} \right)\\
&=\log\{-\log[\mu(\boldsymbol{\theta}_s;A=1)]\}-\log\{-\log[\mu(\boldsymbol{\theta}_s;A=0)]\}. \nonumber
\end{aligned} 
\end{equation}

\noindent The integrals are approximated using the Bayesian bootstrap procedure described in Section \ref{marg_procedures} of the manuscript. After fitting the flexible, semi-parametric proportional hazards model, $s=1,...,S$ samples are obtained from the joint posterior distribution of the model parameters, $\pi(\boldsymbol{\theta} \mid \mathcal{D}_{n_t})$. Let $\boldsymbol{\theta}_s$ represent the $s^{th}$ draw from this joint posterior distribution.  For every row $i = 1,...,{n_t}$ in the sample data, a value of $A_i=1$ is assigned.  Then for each $\boldsymbol{\theta}_s$ the following procedure is performed. For the $t$ corresponding to the time from the start of the trial to the current analysis, the $n_t$ values of the indexed conditional survival probabilities, $\mu(\boldsymbol{\theta}_s;A_i=1,\mathbf{X}_i=\mathbf{x}_i)$ are calculated. A vector $\mathbf{w}_s=(w_{1,s},...,w_{n_t,s})$ is drawn from a Dirichlet($\mathbf{1}_{n_t}$) distribution.  Using $\mathbf{w}_s$, the $n_t$ values are then averaged, $\sum_{i=1}^{n_t}w_{i,s}\mu(\boldsymbol{\theta}_s;A_i=1,\mathbf{X}_i=\mathbf{x}_i)$, which marginalizes them with respect to the observed $\mathbf{X}=\mathbf{x}$, yielding a single sample $\mu(\boldsymbol{\theta}_s;A=1)$ from the posterior distribution of $\mu(\boldsymbol{\theta};A=1)$. For numerical stability, a ${\log\{-\log[\cdot]\}}$ transformation is applied to yield a single sample from the posterior distribution of $\log\{h(t \mid A=1)\}$.  This occurs for all $\boldsymbol{\theta}_s$ to yield $S$ draws from the posterior distribution of ${\log\{h(t \mid A=1)\}}$.  This entire process is then repeated for $A_i=0$, to yield $S$ draws from the posterior distribution of  $\log\{h(t \mid A=0)\}$.  These posterior draws are subtracted and then exponentiated to yield samples from the posterior distribution of the marginal hazard ratio $\gamma(\boldsymbol{\theta})$. A brief summary is below.
\begin{enumerate}
    \item Fit a flexible semi-parametric proportional hazards model.
    \item Obtain $s=1,...,S$ samples from the joint posterior distribution of the model parameters, $\pi(\boldsymbol{\theta} \mid \mathcal{D}_{n_t})$.
    \item \label{tte_lab1} Create one copy of the sample data where $A_i=1$ for all $i=1,...,{n_t}$.
    \item For each $\boldsymbol{\theta}_s$, perform the following: \label{tte_lab2}
    \begin{enumerate}
        \item  For each $i=1,...,{n_t}$, calculate the conditional survival probabilities at time $t$ corresponding to the time from the start of the trial to the current analysis, $\mu(\boldsymbol{\theta}_s;A_i=1,\mathbf{X}_i=\mathbf{x}_i)$.
        \item Sample $\mathbf{w}_s=(w_{1,s},...,w_{n_t,s})$ from a Dirichlet($\mathbf{1}_{n_t}$) distribution.
        \item \label{tte_lab3} Average these $n_t$ values to marginalize with respect to the observed $\mathbf{X}=\mathbf{x}$: \\ $\mu(\boldsymbol{\theta}_s;A=1)=\sum_{i=1}^{n_t}w_i \mu(\boldsymbol{\theta}_s;A_i=1,\mathbf{X}_i=\mathbf{x}_i)$.
        \item \label{tte_lab4} Apply a ${\log\{-\log[\cdot]\}}$ transformation to yield a single sample from the posterior distribution of ${\log\{h(t \mid A=1)\}}$.
    \end{enumerate}
    \item This yields $S$ samples from the posterior distribution of ${\log\{h(t \mid A=1)\}}$.
    \item Repeat steps \ref{tte_lab1}-\ref{tte_lab2} for $A_i=0$ to yield $S$ samples from the posterior distribution of ${\log\{h(t \mid A=0)\}}$.
    \item Subtract and then exponentiate to obtain $S$ samples from the posterior distribution of the marginal hazard ratio, $\gamma(\boldsymbol{\theta})$.
\end{enumerate}

\newpage

\section{Ascertainment of Marginal Estimand Values}\label{ascertainment_marginal_estimands}
\counterwithin{figure}{section}

\subsection{Ascertainment of Marginal Relative Risk (Binary Outcome)}\label{alg:opt_binary}

We first select a value of $\beta_0$ on the log-odds scale in the adjusted data generating models, such that the simulated datasets have a specific marginal control event risk ($p_{ctr}$). We then  use $\beta_0$ and the obtained values of $\phi$ (i.e.,~those where the unadjusted model achieves 50\% and 80\% power, also on the log-odds scale) to select the reported value of the marginal relative risk.

To find $\beta_0$, let $Y$ be a binary outcome.  As a reminder, we define $A$ as the treatment assignment indicator, where $A=1$ means being assigned to the treatment group and $A=0$ means being assigned to the control group. Let $l$ be the number of participants assigned to control, $k$ be the number of participants assigned to treatment, and $l + k = n$ be the total number of participants potentially enrolled in the trial.  Let $\mathbf{X}_{n \times p}$ be the set of covariates used in the adjusted data generating model, and $\mathbf{X}_i$ be the row vector corresponding to the values of the covariates for the $i^{th}$ participant.  Recall the marginal control event risk, $p_{ctr}$, is the risk of having an event in those assigned to control.  Then $p_{ctr}$ can be defined with respect to an adjusted data generating model as follows:
\begin{equation}
    \begin{aligned}
    p_{ctr}&=E[Y|A=0] \\
            &\approx \frac{1}{l}\sum_{i=1}^{l}\hat{E}[Y_i|A_i=0, \mathbf{X}_i] \\
            &=\frac{1}{l}\sum_{i=1}^{l}\text{logit}^{-1}\{\beta_0 + \phi(A_i=0) + \mathbf{X}_i\boldsymbol{\beta}\} \\
            &=\frac{1}{l}\sum_{i=1}^{l}\text{logit}^{-1}\{\beta_0 +  \mathbf{X}_i\boldsymbol{\beta}\} \\
            0 &= \left[\sum_{i=1}^{l}\text{logit}^{-1}\{\beta_0 +  \mathbf{X}_i\boldsymbol{\beta}\} \right] - l \times p_{ctr} \nonumber
    \end{aligned}
\end{equation}

\noindent Given a fixed value of $p_{ctr}$, conditional covariate effects $\boldsymbol{\beta}$ on the log-odds scale, and initial simulation of the treatment assignment and covariate distributions, $\{A, \mathbf{X}\}$, $\beta_0$ can be optimized using the last line above (i.e., using uniroot() in R).  

In the simulations for each relative risk value within each maximum sample size, 5,000 datasets (each with 5,000 participants) were generated using $\{\boldsymbol{\beta}, A, \mathbf{X}\}$ as described under the binary outcome data generating mechanism.  From these, 5,000 values for $\beta_0$ were found and the mean of this distribution was selected as the value of $\beta_0$.  Using this and the value of $\phi$, 5,000 values for the marginal relative risk were obtained by dividing the proportion of events in those assigned to treatment (${\hat{E}[Y|A=1]}$) by the proportion of events in those assigned to control (${\hat{E}[Y|A=0]}$).  The mean of this distribution was then reported as the value of the marginal relative risk corresponding to $\beta_0$ and $\phi$. 

\subsection{Ascertainment of Marginal Hazard Ratio (Time-to-event Outcome)}\label{alg:opt_tte}

Our goal is specify a value of the reported marginal hazard ratio which corresponds to the value of $\phi$ (on the log-hazard scale) used in the adjusted data generating models.  Recall our assumption of proportional hazards, where the marginal hazard ratio is not time-dependent. Let $Y=\{T,\delta\}$ be defined as in the section for hazard ratios in the manuscript.  Define $A$ as the treatment assignment indicator, where $A=1$ means being assigned to the treatment group and $A=0$ means being assigned to the control group. Let $t$ be the maximum duration of the trial and ${P(T>t|A=1)=S(t|A=1)}$ and ${P(T>t|A=0)=S(t|A=0)}$ be the survival probabilities at time $t$ for those assigned to treatment and control, respectively.  In the simulations for each hazard ratio value within each maximum sample size, 5,000 datasets (each with 5,000 participants) were generated using ${\{\boldsymbol{\beta}, A, \mathbf{X}, t = 50\}}$ as described under the time-to-event outcome data generating mechanism.  For each dataset, the value of the marginal hazard ratio was calculated as:
\begin{equation}
    \begin{aligned}
        \gamma=\exp\{\log(-\log[\hat{P}(T>50|A=1)]) - \log(-\log[\hat{P}(T>50|A=0)])\} \nonumber
    \end{aligned}
\end{equation}

\noindent The mean of this distribution was then reported as the value of the marginal hazard ratio corresponding to $\phi$. 

\newpage

\section{CCEDRN COVID-19 RCT Truncated Covariate Distributions}\label{alg:trunc_cov}
\counterwithin{figure}{section}

The following algorithm yields datasets from truncated Normal distribution $F$ with correct post-truncation minimum $\wedge$, maximum $\vee$, $q1$, and $q3$ values and approximately correct post-truncation $\mu$ and $\sigma$ values.
\begin{enumerate}
    \item Determine maximum sample size of trial, $max\_ss$.
    \item Obtain reported summary statistics $\{q1,q3,\mu,\sigma\}$ and range $[\wedge,\vee]$.
    \item Set plausible value of median $q2$ if not reported.
    \item Set $n=250 \times max\_ss$.
    \item Select starting values for $\{\xi,\tau^2\}$.
    \item Until $\mu^* \approx \mu$ and $\sigma^* \approx \sigma$:
    \begin{enumerate}
        \item Generate $i=1,...,n$ values of $X_i$ from $N(\xi,\tau^2)$.
        \item Discard $X_i \not\in [min,max]$.
        \item Sample $\frac{n}{4}$ values of $X_i \in [\wedge, q1]$.
        \item Sample $\frac{n}{4}$ values of $X_i \in [q1, q2]$.
        \item Sample $\frac{n}{4}$ values of $X_i \in [q2, q3]$.
        \item Sample $\frac{n}{4}$ values of $X_i \in [q3, \vee]$.
        \item Collect all and set $\mu^*$ and $\sigma^*$ as the mean and standard deviation of the sampled values. 
        \item Update $\{\xi,\tau^2\}$ or break if $\mu^* \approx \mu$ and $\sigma^* \approx \sigma$.
    \end{enumerate}
    \item To generate one dataset from $F$, use final values of $\{\xi,\tau^2\}$ to repeat process above, but sample $max\_ss$ values from the $n$ collected values.
\end{enumerate}

\noindent For age, the following summary statistics and simulation parameter values were used:
\begin{equation}
    \{\wedge=18, q1=39, q2=55, q3=70, \vee=90, \mu=54.7, \sigma=19.8, \xi=62, \tau=40\}. \nonumber
\end{equation}

\noindent For respiratory rate, the following summary statistics and simulation parameter values were used:
\begin{equation}
    \{\wedge=12, q1=18, q2=20, q3=22, \vee=40, \mu=21.0, \sigma=6.2, \xi=30, \tau=6\}. \nonumber
\end{equation}

\newpage

\section{Summary graphics for bias and RMSE}
\counterwithin{figure}{section}

\begin{figure}[h]
    \includegraphics[width=1\textwidth]{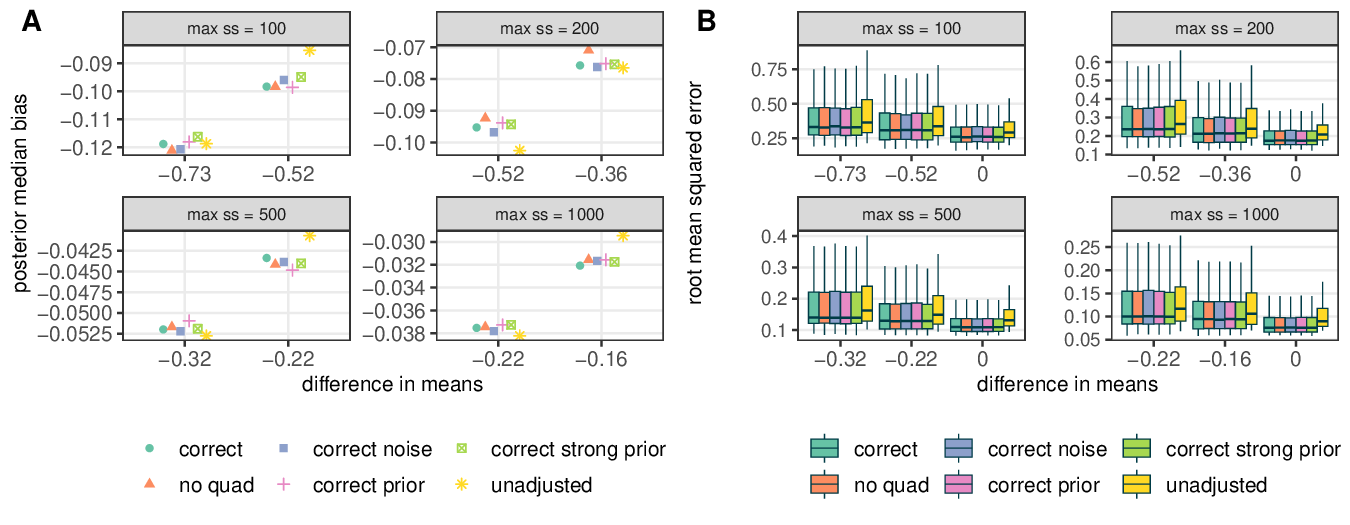}
    \caption{Continuous outcome. A) Posterior median bias and B) root mean squared error. Panels correspond to various maximum sample sizes (max ss). Points are jittered horizontally.}
    \centering
    \label{fig:cont_bias_rmse}
\end{figure}

\begin{figure}[h]
    \includegraphics[width=1\textwidth]{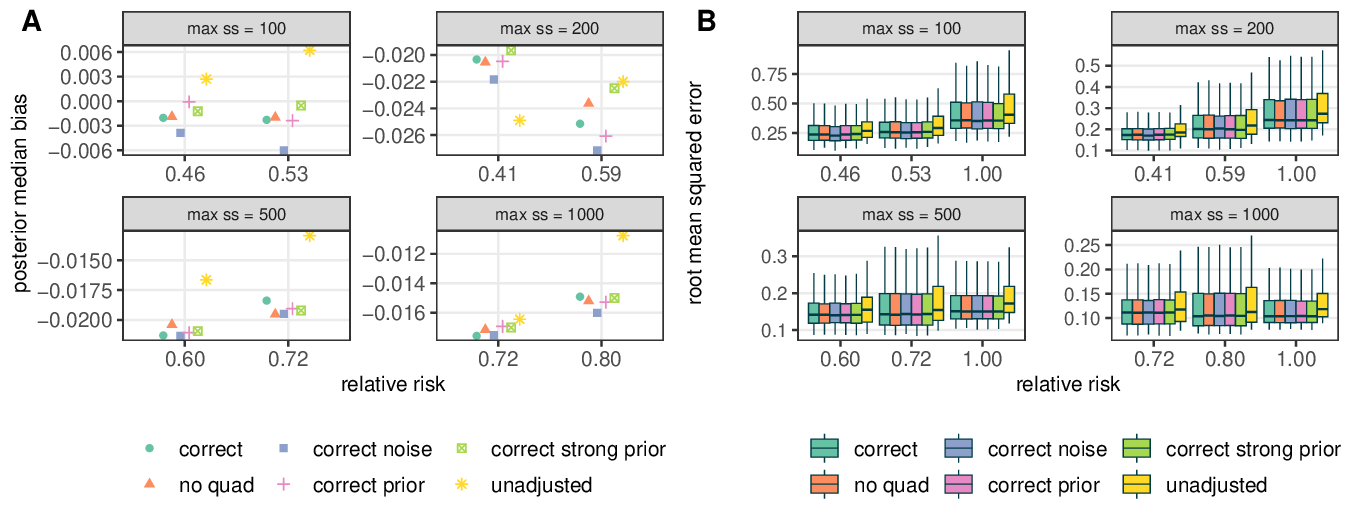}
    \caption{Binary outcome. A) Posterior median bias and B) root mean squared error. Panels correspond to various maximum sample sizes (max ss). Points are jittered horizontally.}
    \centering
    \label{fig:binary_bias_rmse}
\end{figure}

\begin{figure}[h]
    \includegraphics[width=1\textwidth]{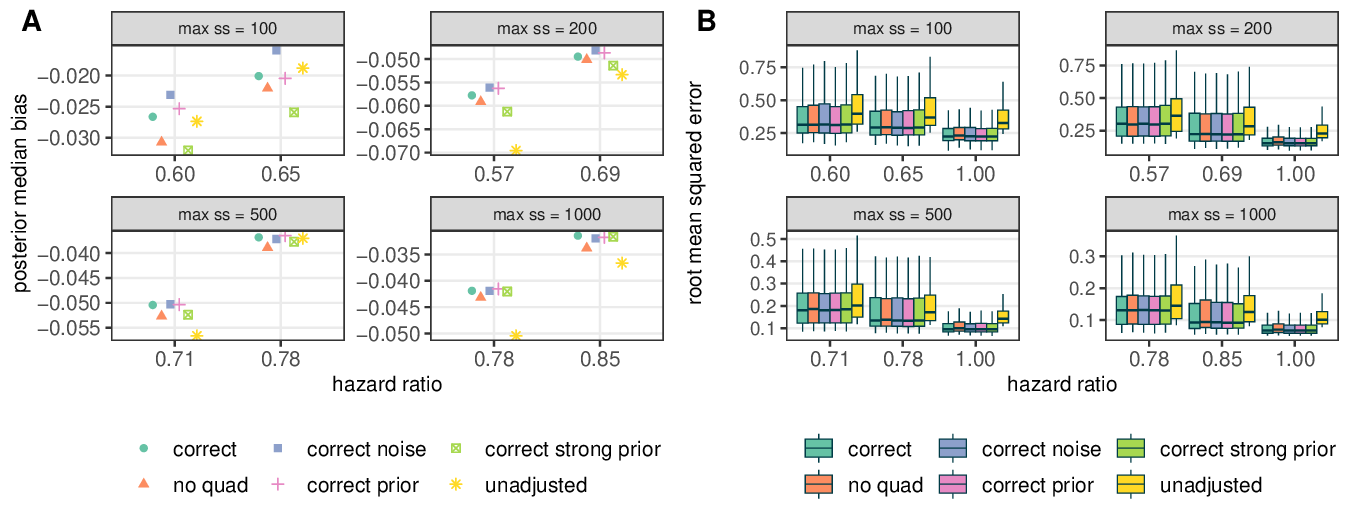}
    \caption{Time-to-event outcome. A) Posterior median bias and B) root mean squared error. Panels correspond to various maximum sample sizes (max ss). Points are jittered horizontally.}
    \centering
    \label{fig:tte_bias_rmse}
\end{figure}

\begin{figure}[h]
    \centering
    \includegraphics[width=0.7\textwidth]{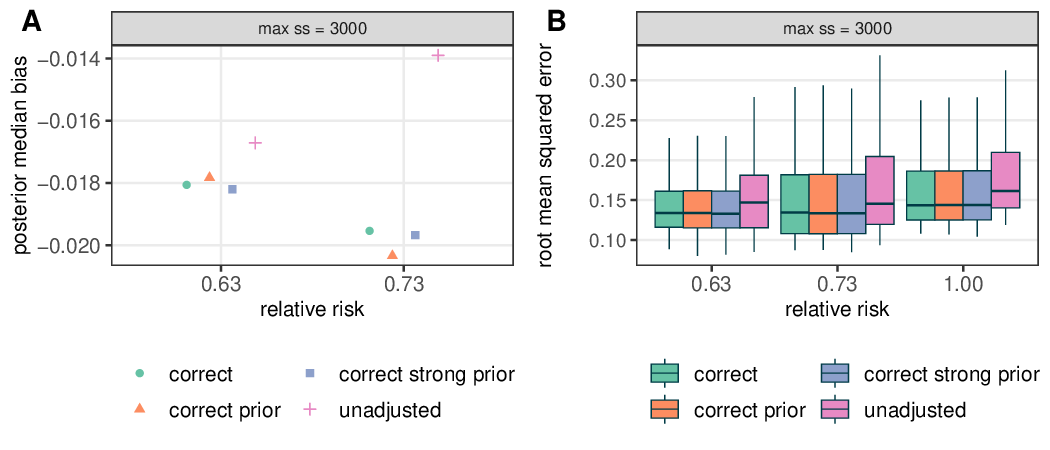}
    \caption{CCEDRN-ADAPT COVID-19 trial with binary outcome. A) Posterior median bias and B) root mean squared error. Panels correspond to a maximum sample size (max ss) of 3,000. Points are jittered horizontally.}
    \label{fig:covid_bias_rmse}
\end{figure}

\clearpage

\section{Simulation for Informative Prior on Treatment Effect}
\counterwithin{figure}{section}
\counterwithin{table}{section}
\begin{figure}[ht]
    \centering
     \includegraphics[width=1\textwidth]{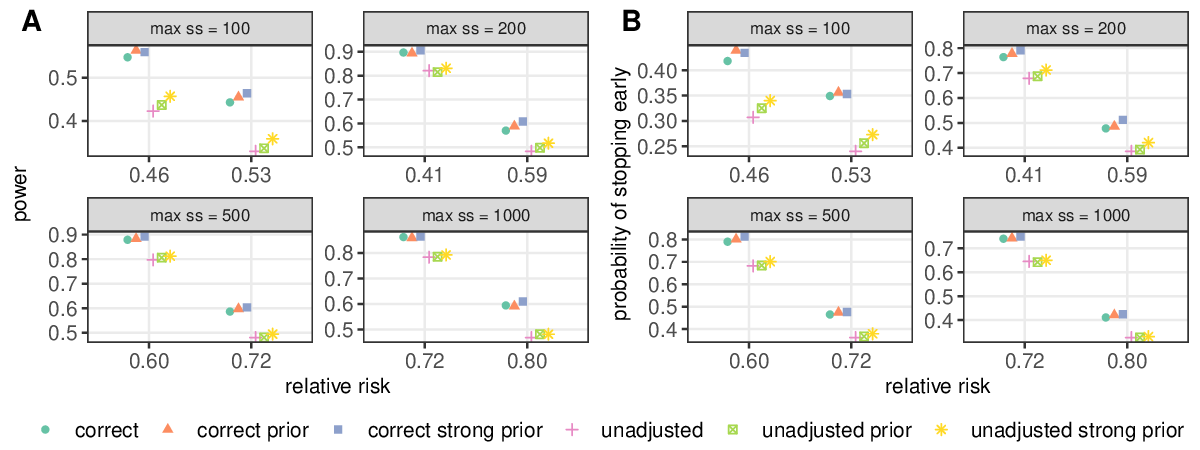}
    \caption{Binary outcome. A) Power and B) probability of stopping early. Panels correspond to various maximum sample sizes (max ss). Points are jittered horizontally.}
    \label{fig:inf_prior_trt_effect}
\end{figure}

In this section, we consider the impact of incorporating informative prior information on the treatment effect for trials with binary endpoints. Six adjustment models are considered:
\begin{enumerate}
    \item correct: $\beta_0 + \phi A+\beta_1X_1 +\beta_2X_2 +\beta_3X_3 +\beta_4X_3^2 +\beta_5X_5$ 
    \item correct prior: $\beta_0 + \phi A+\beta_1^{\dagger}X_1 +\beta_2^{\dagger}X_2 +\beta_3^{\dagger}X_3 +\beta_4^{\dagger}X_3^2 +\beta_5^{\dagger}X_5$ 
    \item correct strong prior: $\beta_0 + \phi A+\beta_1^{\dagger\dagger}X_1 +\beta_2^{\dagger\dagger}X_2 +\beta_3^{\dagger\dagger}X_3 +\beta_4^{\dagger\dagger}X_3^2 +\beta_5^{\dagger\dagger}X_5$ 
    \item unadjusted: $\beta_0 +  \phi A$
    \item unadjusted prior: $\beta_0 + \phi^{\dagger} A$ 
    \item unadjusted strong prior: $\beta_0 + \phi^{\dagger\dagger} A$ 
\end{enumerate}

\noindent The regression coefficients $\{\phi, \boldsymbol{\beta}, \boldsymbol{\beta}^{\dagger},\boldsymbol{\beta}^{\dagger\dagger}\}$ and the functional forms of the \textit{correct}, \textit{correct prior}, \textit{correct strong prior}, and \textit{unadjusted} models are defined as previously described for the binary trials in the simulation study section of the manuscript. The \textit{unadjusted prior} model includes a prior on the treatment indicator coefficient centered at the value used in the data generating mechanism where the \textit{unadjusted} model achieves approximately 80\% power (50\% power for max ss = 100).  The \textit{unadjusted strong prior} model both centers and re-scales this prior to be more informative. The following priors are then used for the treatment indicator coefficients in the \textit{unadjusted prior} and \textit{unadjusted strong prior} models:
\begin{equation}
\begin{aligned}
    \phi^{\dagger} &\sim \text{Normal}(c, 2.5/s_a) \\
    \phi^{\dagger\dagger} &\sim \text{Normal}(c, 1/s_a) \nonumber
\end{aligned}
\end{equation}
where $c=\{-1.21, -1.36, -0.82, -0.54\}$ for $\text{max ss} = \{100,200,500,1000\}$. All other components for the binary trials remain as described in the simulation study section of the manuscript.

Results for power and the probability of stopping early are displayed in Figure \ref{fig:inf_prior_trt_effect}. Including stronger priors on the treatment effect may increase the power and probability of stopping early as compared to weakly informative priors.  This hold for both the \textit{correct} and \textit{unadjusted} model variants and is most beneficial for smaller sample sizes.  However, this comes at the cost of inflated type 1 error (T1E), with the greatest inflation occurring for the smaller maximum sample sizes (Table \ref{tbl:inf_prior_trt_effect}). Both the type 1 error inflation and increase in power become less pronounced in the trials with larger maximum sample sizes where the priors are dominated by the data.

\begin{table}[th]
\centering
\caption{Binary outcome. Type 1 error rate (T1E), bias under the null ($\text{Bias}^*$), and expected sample size at three different values of the marginal relative risk ($\gamma$).}
\resizebox{\linewidth}{!}{
\begin{tabular}[t]{lrrrrrrrrrr}
\toprule
\multicolumn{1}{c}{ } & \multicolumn{5}{c}{Maximum sample size = 100} & \multicolumn{5}{c}{Maximum sample size = 200} \\
\cmidrule(l{3pt}r{3pt}){2-6} 
\cmidrule(l{3pt}r{3pt}){7-11} 
\multicolumn{3}{c}{ } & \multicolumn{3}{c}{Expected sample size} & \multicolumn{2}{c}{ } & \multicolumn{3}{c}{Expected sample size} \\
\cmidrule(l{3pt}r{3pt}){4-6} \cmidrule(l{3pt}r{3pt}){9-11}
Adjustment model & T1E & $\text{Bias}^*$ & $\gamma=1$ & $\gamma=0.53$ & $\gamma=0.46$ & T1E & $\text{Bias}^*$ & $\gamma=1$ & $\gamma=0.59$ & $\gamma=0.41$\\
\midrule
correct & 0.063 & 0.031 & 97.0 & 86.3 & 83.6 & 0.036 & 0.021 & 196.8 & 162.5 & 138.3\\
correct prior & 0.071 & 0.021 & 96.5 & 85.7 & 82.2 & 0.038 & 0.017 & 196.5 & 161.9 & 136.5\\
correct strong prior & 0.063 & -0.014 & 97.3 & 86.6 & 83.8 & 0.046 & -0.006 & 195.7 & 159.2 & 134.9\\
unadjusted & 0.034 & 0.058 & 98.6 & 90.7 & 88.3 & 0.031 & 0.025 & 198.0 & 171.4 & 147.2\\
unadjusted prior & 0.034 & 0.051 & 98.6 & 90.0 & 87.4 & 0.032 & 0.021 & 197.9 & 170.8 & 146.2\\
unadjusted strong prior & 0.041 & 0.009 & 98.4 & 89.6 & 86.8 & 0.034 & 0.001 & 197.6 & 167.6 & 142.8\\
\midrule
\multicolumn{1}{c}{ } & \multicolumn{5}{c}{Maximum sample size = 500} & \multicolumn{5}{c}{Maximum sample size = 1000} \\
\cmidrule(l{3pt}r{3pt}){2-6} 
\cmidrule(l{3pt}r{3pt}){7-11} 
\multicolumn{3}{c}{ } & \multicolumn{3}{c}{Expected sample size} & \multicolumn{2}{c}{ } & \multicolumn{3}{c}{Expected sample size} \\
\cmidrule(l{3pt}r{3pt}){4-6} \cmidrule(l{3pt}r{3pt}){9-11}
Adjustment model & T1E & $\text{Bias}^*$ & $\gamma=1$ & $\gamma=0.72$ & $\gamma=0.60$ & T1E & $\text{Bias}^*$ & $\gamma=1$ & $\gamma=0.80$ & $\gamma=0.72$\\
\midrule
correct & 0.028 & 0.010 & 493.7 & 404.9 & 334.9 & 0.022 & 0.008 & 992.0 & 823.2 & 681.5\\
correct prior & 0.028 & 0.009 & 494.1 & 402.9 & 334.9 & 0.021 & 0.007 & 991.7 & 818.9 & 680.3\\
correct strong prior & 0.028 & 0.004 & 493.7 & 401.2 & 329.9 & 0.023 & 0.005 & 990.5 & 813.9 & 675.7\\
unadjusted & 0.026 & 0.016 & 494.6 & 426.0 & 367.0 & 0.024 & 0.010 & 990.2 & 859.6 & 734.2\\
unadjusted prior & 0.029 & 0.015 & 493.4 & 426.7 & 364.9 & 0.024 & 0.009 & 989.5 & 858.6 & 733.7\\
unadjusted strong prior & 0.031 & 0.009 & 492.6 & 423.0 & 362.7 & 0.023 & 0.007 & 989.5 & 858.1 & 733.9\\
\bottomrule
\end{tabular}}
\label{tbl:inf_prior_trt_effect}
\end{table}

\clearpage

\section{Bias From Overestimation in Trials which Stop Early for Superiority}
\counterwithin{figure}{section}

\begin{figure}[h]
    \centering
    \includegraphics[width=1\textwidth]{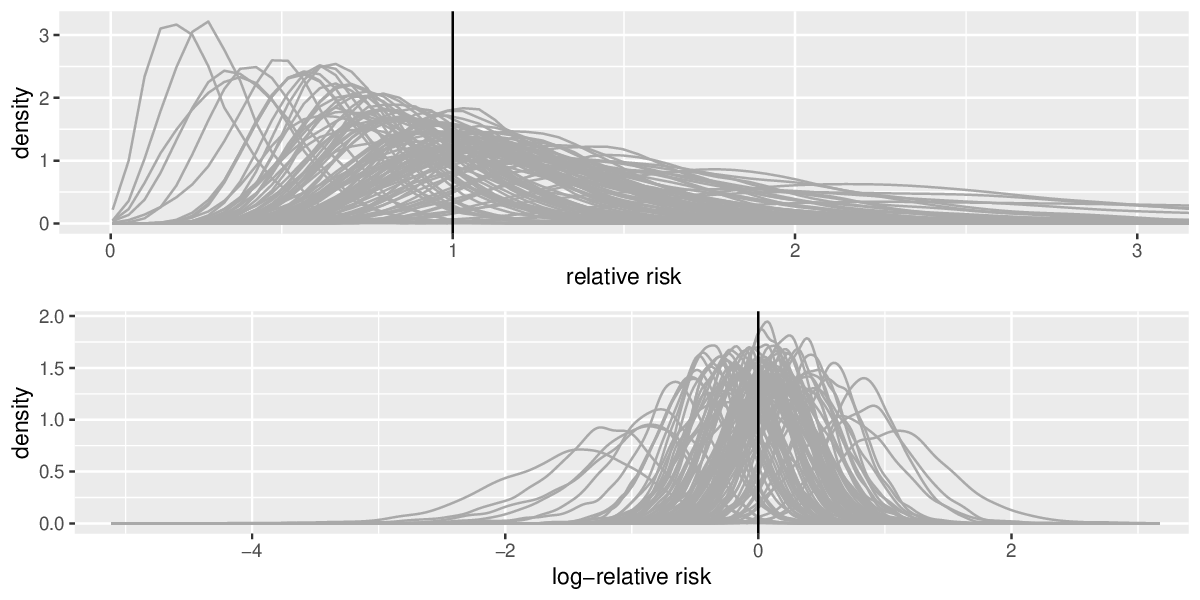}
    \caption{100 posterior distributions of the relative risk (top) and log-relative risk (bottom) for a trial with a binary endpoint and maximum sample size of 100. Vertical line represents the null treatment effect.}
    \label{fig:post_samps_bias}
\end{figure}

\cite{walter2019randomised} shows that overestimation is to be expected when trials permit early stopping for superiority. They consider the case of frequentist group-sequential designs and compare three different stopping rules which differ in how the overall $\alpha$ is divided among the interim and final analyses. The Pocock, O’Brien, and Fleming (PCK) stopping rule evenly divides $\alpha$ across all analyses (interim and final) keeping the stringency of the stopping criteria constant. This is the frequentist group sequential stopping rule most like the Bayesian stopping rule employed in the current manuscript, where a single value for the upper probability threshold u is used (u=0.99), thereby also keeping the stringency of the stopping rule constant across all interim and final analyses.  It is shown that overestimation is expected for the PCK stopping rule, and so we conclude it should also be expected for the Bayesian stopping rule employed here, thus inducing the observed bias in the treatment effect under the simulation scenarios. In Section 3.1 of \cite{walter2019randomised}, the authors discuss observing greater over-estimation for the Haybittle and Peto (HP) stopping rule as compared to the PCK stopping rule. They state ``rules (such as HP) that have a more stringent threshold for stopping involve a greater risk of over-estimation if the rule is actually invoked." This suggests that Bayesian stopping rules which have more stringent initial stopping criteria may lead to increased bias as compared to the stopping rules employed in the current manuscript, though we do not explore this further here. 

\begin{figure}
    \centering
    \includegraphics[width=1\textwidth]{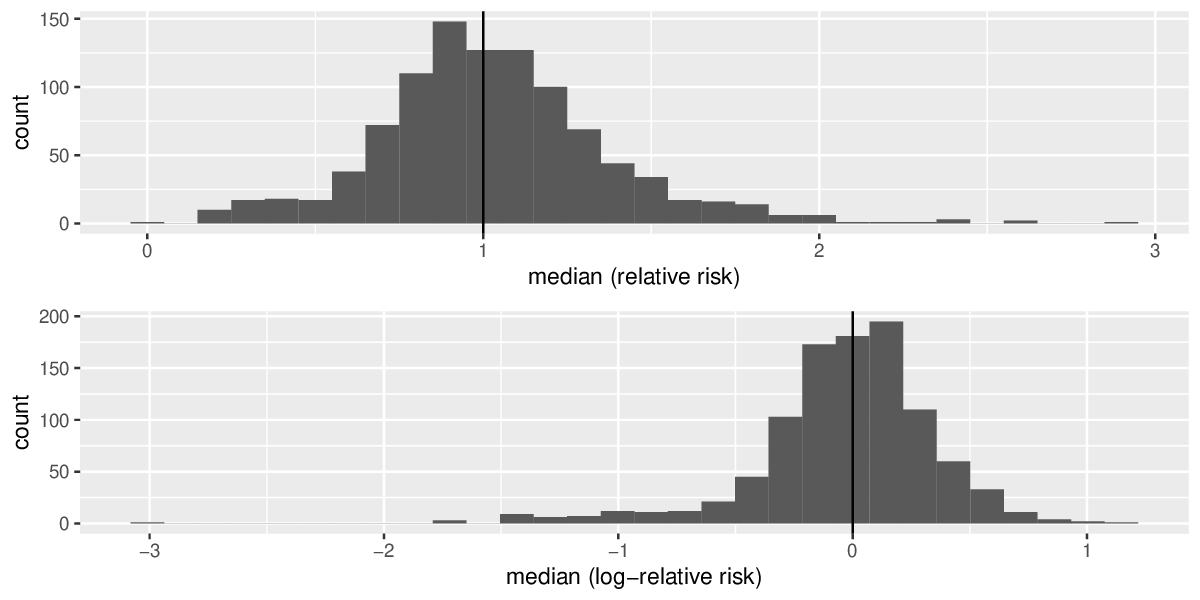}
    \caption{100 posterior medians of the relative risk (top) and log-relative risk (bottom) for a trial with a binary endpoint and maximum sample size of 100. Vertical line represents the null treatment effect.}
    \label{fig:post_meds_bias}
\end{figure}

Considering bias under the null, the difference in signs between the continuous endpoint versus the binary and time-to-event endpoints reflects the lower bounds of the estimands.  The difference in means under the continuous endpoint is unbounded below, whereas the relative risk and hazard ratios are bounded below by zero.  When bias is calculated on the log-relative risk and log-hazard ratio scales, most values for bias under the null for both the binary and time-to-event endpoints become negative as well, with greater bias for smaller sample sizes as in the continuous endpoint. This is explained visually in Figure \ref{fig:post_samps_bias} and \ref{fig:post_meds_bias}, where 100 posterior distributions (Figure \ref{fig:post_samps_bias}) and posterior medians (Figure \ref{fig:post_meds_bias}) have been plotted for the null treatment effect for a trial with a binary endpoint under a maximum sample size of 100. The vertical lines represent the null values used for calculation of bias.

In the top panel of Figure \ref{fig:post_samps_bias} on the relative risk scale, we see many right-skewed posteriors which lead to some posterior median estimates which are much greater than the null (Figure \ref{fig:post_meds_bias}, top panel). This induces positive bias under the null.  When we move to the log scale, the right-skewed distributions become more normal in shape and are more centered around the null, but some of the posteriors which were closer to the lower boundary of 0 on the relative risk scale become left-skewed (Figure \ref{fig:post_samps_bias}, bottom panel).  This results in some posterior medians becoming much less than the null (Figure \ref{fig:post_meds_bias}, bottom panel) which leads to negative bias under the null as in the continuous endpoint case.  When calculating bias for non-null treatment effects on the relative risk and hazard ratio scales, the posteriors are pushed further toward 0 than in the figures included in this section which is why they can still attain negative values and clearly exhibit overestimation in these cases. 

\clearpage
\section{Example of the Non-Collapsibility of the Odds Ratio}
\counterwithin{figure}{section}

\begin{figure}[h]
    \centering
    \includegraphics[width=0.65\textwidth]{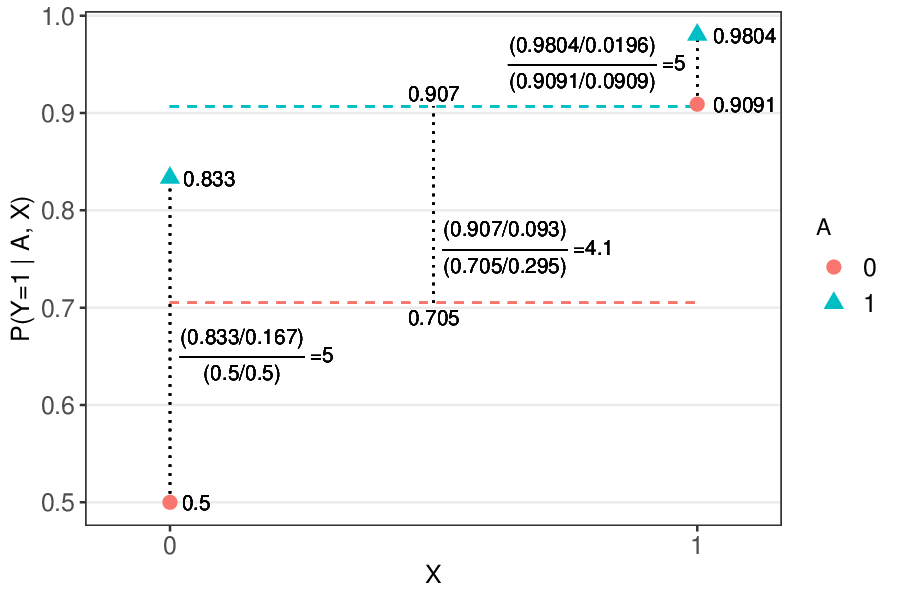}
    \caption{Example of non-collapsibility of the odds ratio.  Colored circles and triangles correspond to values of risk under treatment or control for different values of $X$.}
    \label{fig:non_collapsibility}
\end{figure}
We consider a slightly modified version of the example provided in \cite{daniel2021making}. Consider a RCT with a binary endpoint following the logistic regression model below which contains a binary covariate $X$ and binary treatment indicator $A$, and where $P(X=1)=P(A=1)=0.5$. Define $\phi = \log(5)$, $\beta = \log(10)$, and $\boldsymbol{\theta}=\{\phi,\beta\}$. We assume the following model:
\begin{equation}
    \text{logit}(P(Y=1 \mid A, X)) = \phi A + \beta X. \nonumber
\end{equation}
In this example, the conditional odds ratio for those who are treated versus untreated is 5, regardless of the value of $X$. To see this, define $\mu(\boldsymbol{\theta}; A, X) = P(Y=1 \mid A, X) = \text{logit}^{-1}(\phi A + \beta X)$. For $X=0$, we have $\mu(\boldsymbol{\theta}; A=1, X=0)=0.833$ and that $1 - \mu(\boldsymbol{\theta}; A=1, X=0)=0.167$ yielding the odds for the event in those who are treated to be $\mu(\boldsymbol{\theta}; A=1, X=0) / 1 - \mu(\boldsymbol{\theta}; A=1, X=0) = 0.833/0.167 = 5$.  For those who are untreated, we have $\mu(\boldsymbol{\theta}; A=0, X=0) = 0.5$ and that $1 - \mu(\boldsymbol{\theta}; A=0, X=0)=0.5$ yielding odds of $\mu(\boldsymbol{\theta}; A=0, X=0) / 1 - \mu(\boldsymbol{\theta}; A=0, X=0) = 0.5/0.5 = 1$. Dividing these yields a conditional odds ratio for those who are treated versus untreated under $X=0$ to be $5/1=5$. Similar calculations for $X=1$ yield $\mu(\boldsymbol{\theta}; A=1, X=1)=0.9804$, $1 - \mu(\boldsymbol{\theta}; A=1, X=1)=0.0196$ yielding the odds for the event in those who are treated to be $0.9804/0.0196 = 50$. For those who are untreated, we have $\mu(\boldsymbol{\theta}; A=0, X=1) = 0.9091$ and that $1 - \mu(\boldsymbol{\theta}; A=0, X=1)=0.0909$ yielding odds of $0.9091/0.0909 = 10$. Dividing these yields a conditional odds ratio for those who are treated versus untreated under $X=1$ to be $50/10=5$. In Figure \ref{fig:non_collapsibility}, we that these conditional odds ratios correspond to a vertical comparison of the risks under the treatment assignments $A$ for either value of $X$ (dotted vertical lines). To obtain the marginal odds ratio, we must average these risks with respect to the distribution of $X$. This yields the horizontal dashed lines (colored by value of $A$) where $\mu(\boldsymbol{\theta}; A = 1)=0.5(0.833) + 0.5(0.980) = 0.907$ and $\mu(\boldsymbol{\theta}; A = 0)=0.5(0.5) + 0.5(0.909) = 0.705$. The marginal odds ratio then corresponds to a vertical comparison of these horizontal lines.  Doing so yields a marginal odds ratio of $\gamma(\boldsymbol{\theta})=(0.907/0.093)/(0.705/0.295)=4.1$. We see that the marginal odds ratio is not equal to the conditional odds ratio, and thus the odds ratio is non-collapsible.

This same example can be viewed using two-by-two tables, where the cells contain the proportions expected under each combination of treatment assignment and covariate value.

\begin{table}[h]
    \centering
    \begin{tabular}{c|cc|c}
\multicolumn{1}{c}{} & \multicolumn{2}{c}{X = 0} & \multicolumn{1}{c}{}  \\
& $A=1$ & $A=0$ & $P(Y=y)$ \\
 \hline
$Y=1$ & 0.8333333 & 0.5 & 0.667 \\
$Y=0$ & 0.16666667 & 0.5 & 0.333 \\
 \hline
$P(A=a)$ & 0.5 & 0.5 & 1 \\
    \end{tabular}
\quad
\begin{tabular}{c|cc|c}
\multicolumn{1}{c}{} & \multicolumn{2}{c}{X = 1} & \multicolumn{1}{c}{} \\
& $A=1$ & $A=0$ & $P(Y=y)$ \\
 \hline
$Y=1$ & 0.9803922 & 0.9090909 & 0.945 \\
$Y=0$ & 0.01960784 & 0.09090909 & 0.055 \\
 \hline
$P(A=a)$ & 0.5 & 0.5 & 1 \\
    \end{tabular}
    \label{tab:cond_or}
\end{table}

The conditional odds ratio from each table is 5. Below, we consider the marginal table with proportions expected under each combination of treatment assignment and covariate value. This is found by averaging the risk values in the conditional tables with respect to the distribution of $X$, where we recall that $P(X=1)=0.5$.

\begin{table}[ht]
    \centering
    \begin{tabular}{c|cc|c}
& $A=1$ & $A=0$ & $P(Y=y)$ \\
 \hline
$Y=1$ & 0.90686275 & 0.70454545 & 0.806 \\
$Y=0$ & 0.09313725 & 0.29545455 & 0.194 \\
 \hline
$P(A=a)$ & 0.5 & 0.5 & 1 \\
    \end{tabular}
    \label{tab:marg_or}
\end{table}

The marginal odds ratio from the table above is 4.1. We observe that the true conditional and marginal odds ratios are not equal, thus showing the odds ratio is non-collapsible.

\end{appendices}

\end{singlespace}
\end{document}